\documentclass{style}
\usepackage{graphicx}
\usepackage{amsmath}
\usepackage[ulem=normalem]{changes} 

\begin{document}

\title{Symmetries and deformations in the spherical shell model}
\author{
P~Van~Isacker\email{isacker@ganil.fr} \\
  \it Grand Acc\'el\'erateur National d'Ions Lourds, CEA/DSM--CNRS/IN2P3\\
  \it BP~55027, F-14076 Caen Cedex 5, France\\
S~Pittel\\
  \it Bartol Research Institute and Department of Physics and Astronomy\\
  \it University of Delaware, Newark, DE 19716, USA}


\date{}

\maketitle

\begin{abstract}
We discuss symmetries of the spherical shell model
that make contact with the geometric collective model of Bohr and Mottelson.
The most celebrated symmetry of this kind is SU(3),
which is the basis of Elliott's model of rotation.
It corresponds to a deformed mean field
induced by a quadrupole interaction in a single major oscillator shell $N$
and can be generalized to include several major shells.
As such, Elliott's SU(3) model establishes the link between the spherical shell model
and the (quadrupole component of the) geometric collective model.
We introduce the analogue symmetry
induced by an octupole interaction in two major oscillator shells $N-1$ and $N$,
leading to an octupole-deformed solution of the spherical shell model.
We show that in the limit of large oscillator shells, $N\rightarrow\infty$,
the algebraic octupole interaction tends to that of the geometric collective model.
\end{abstract}

\section{Introduction}
\label{s_intro}
Our understanding of the structure of the atomic nucleus is at present incomplete,
based as it is on various models with limited ranges of applicability.
Examples are the spherical shell model and the geometric collective model.
While the former stresses the single-particle nature of the nucleons in the nucleus,
their coherent motion is emphasized by the latter.
A recurring question in more than half a century of basic nuclear research has been
how to reconcile two such opposing views of the structure of the nucleus.
The currently accepted paradigm is
that the structural patterns and regularities,
as predicted by the geometric collective model,
arise as emergent behaviour from the complex many-body problem
of nucleonic interactions in the context of the spherical shell model.

While this connection can, perhaps, be established in principle,
and much current work is going on along these lines,
many of its aspects still remain unclear.
In this paper we study this question from the perspective of symmetries.
Ever since the pioneering studies by Wigner, Racah and Elliott,
symmetry considerations have played a pivotal role
in the development of nuclear models,
in particular of the spherical shell model.
We do not intend to review here all such symmetries
but focus on those that make contact
with the geometric collective model.

An alternative microscopic approach to nuclear collective motion
that is under intense current investigation
involves the use of self-consistent mean-field models~\cite{RS80},
whereby nuclei are described variationally
from realistic interactions between their constituent nucleons.
There has also been much recent work
to take these models beyond mean field~\cite{Bender03}.
Models based on symmetries provide exactly solvable limits
for the same collective features described by these mean-field approaches,
but at the cost of using semi-realistic, rather than fully realistic, hamiltonians.
Nevertheless, because of the exact solvability of these symmetry-based models,
unique perspectives on collective properties can often be identified.

We start by painting a qualitative picture of the two models
we aim connect in this paper in sections~(\ref{ss_sm}) and~(\ref{ss_bm}).
One symmetry of the spherical shell model is SU(3),
which forms the basis of our understanding of nuclear rotation.
It can be given a rigorous formulation
in the context of Wigner's supermultiplet model,
which is reviewed in section~\ref{s_wigner}.
Elliott's SU(3) model of quadrupole deformation
is discussed in section~\ref{s_elliott},
first in its elementary version applied to a single major shell of the harmonic oscillator
and subsequently in its extended version applied to several shells.
Section~\ref{s_octu} deals with the analogous problem of octupole deformation 
and proposes a corresponding symmetry in the context of the spherical shell model.
Some concluding remarks are made in section~\ref{s_conc}.

\subsection{A shell-model primer}
\label{ss_sm}
In first approximation the structure of a nucleus
is determined by the nuclear mean field,
that is, the average potential felt by each nucleon
resulting from the interactions with all others.
In particular, the observed shell structure of nuclei
can be understood on the basis of the notion of mean field.
However, for a detailed description of many nuclear properties,
a residual nucleon--nucleon interaction on top of the average potential
must be taken into account.
Nuclear structure at low energies
is particularly affected by the residual interaction
between nucleons in the valence shell,
that is, in the outermost shell that is not completely filled.

While a current focus of nuclear structure theory
aims at a consistent microscopic derivation
of the nuclear mean field and the residual interaction,
several of the basic features of the structure of nuclei
are captured in the following schematic nuclear hamiltonian:
\begin{align}
\hat H={}&
\sum_{k=1}^A
\left({\frac{p_k^2}{2m_{\rm n}}}+
{\frac 1 2}m_{\rm n}\omega^2r_k^2+
\zeta_{\ell\ell}\ell_k^2+
\zeta_{\ell s}\bar \ell_k\cdot\bar s_k\right)
\nonumber\\&+
\sum_{1\leq k<l}^A
\hat V_{\rm ri}(\xi_k,\xi_l),
\label{e_hamsm}
\end{align}
where $A$ is the atomic mass number
(the number of nucleons in the nucleus),
$\xi_k$ is a short-hand notation
for the spatial coordinates,
the spin and isospin variables of nucleon $k$,
$\xi_k\equiv\{\bar r_k,\bar s_k,\bar t_k\}$,
and $m_{\rm n}$ is the nucleon mass.
The first term in equation~(\ref{e_hamsm})
is the kinetic energy of the nucleons.
The second term is a harmonic-oscillator potential
with frequency $\omega$,
which is governed by the nuclear size
and which is a crude approximation
to the nuclear mean field~\cite{Bohr69} for well-bound nuclei.
A more realistic nuclear mean field,
({\it e.g.}, a Woods--Saxon potential)
does not display the degeneracy
of states with different orbital angular momentum $\ell$
in the same major shell,
characteristic of a harmonic oscillator.
To some extent this deficiency of the harmonic-oscillator potential
can be remedied by adding an $\ell^2$ orbit--orbit term,
which lifts the $\ell$-degeneracy
and gives rise to a single-particle spectrum more in accord with that from experiment.
The fourth term in the hamiltonian~(\ref{e_hamsm})
corresponds to a spin--orbit coupling in the nucleonic motion,
whose assumption was the decisive step in the justification
of the nuclear shell model~\cite{Mayer49,Jensen49}
by providing a natural explanation
of the observed `magic' numbers,
those neutron $N$ or proton $Z$ numbers
for which the nucleus acquires an increased stability.
The fact that the spin--orbit interaction is strong
gives rise to another very important feature in nuclear structure,
namely that the higher major shells contain orbitals
that intrude from the next oscillator shell
and have the opposite parity from the others in the major shell.
The last term in equation~(\ref{e_hamsm}) is the residual two-body interaction.
It depends in a complex fashion on the mean field
and on the valence space that is made available to the nucleons.
Because of its dependence on the space in which it acts,
it is often referred to as an {\em effective} interaction.

If single-particle energy spacings are large
compared to a typical matrix element of the residual interaction,
nucleons move independently.
This limit corresponds to the shell model of independent particles.
The neglect of the last term in the hamiltonian~(\ref{e_hamsm})
leads to uncorrelated many-particle eigenstates
that are Slater determinants
constructed from the single-particle eigenfunctions of the harmonic oscillator.
Slater determinants involve products of single-particle states,
organized so that the wave function is fully anti-symmetric under particle interchange,
as required by the Pauli Principle.
This is the key to the shell structure exhibited by nuclei.
If the residual interaction is not neglected,
a true many-body problem results,
where it is critical to incorporate configuration mixing of the Slater determinants
resulting from the residual interaction.
Usually it is only necessary to include configuration mixing
within a single major shell for neutrons and a single major shell for protons,
except fairly near shell closure,
where coherent excitations from other major shells
can be lowered into the region of low-lying states of the dominant shell
and must therefore be considered on the same footing.

A good approximation to the residual effective interaction
for use in a shell-model treatment of nuclei
involves a pairing interaction $\hat V_{\rm pairing}$,
acting between pairs of alike nucleons in time-reversed orbits,
and a sum of separable two-body interactions
acting between all nucleons, {\it viz.}
\begin{equation}
\hat V_{\rm ri}=\hat V_{\rm pairing}+\sum_J \alpha_J\hat P^\dag_J\cdot\hat P_J,
\label{e_vres}
\end{equation}
where $\hat P^\dag_J$ creates a coherent particle--hole pair with multipolarity $J$ and parity $(-)^J$.
In fairly light nuclei the connection between modern realistic effective interactions
and a schematic sum over separable interactions has been carefully demonstrated~\cite{Caurier05}.

The term with $J=0$, called the monopole interaction,
has the primary effect of evolving the single-particle energies of the nuclear mean field,
sometimes even changing the order of single-particle levels
and the magic numbers~\cite{Caurier05}.
The other terms govern the mixing of simple shell-model configurations
that give rise to the wide variety of features seen in nuclei across the periodic table.

The pairing component of the residual interaction,
$\hat V_{\rm pairing}$ in equation~(\ref{e_vres}), has far-reaching consequences.
Perhaps most importantly, it gives rise to pairing correlations,
which have a pervasive impact on nuclear structure properties
throughout the periodic table.
For the purposes of this discussion, however, the pairing interaction,
as well as the related delta interaction, $\delta(\vec{r}_1-\vec{r}_2)$, 
conserve total orbital angular momentum $L$ and total spin $S$,
besides total angular momentum $J$ associated with rotational invariance. 
When this term dominates,
it leads to a classification called $LS$ (or Russell--Saunders) coupling.
This is strongly broken, however, by the spin--orbit term in the nuclear mean field,
which favours $jj$ coupling.
The conflict between $LS$ coupling and $jj$ coupling
plays a crucial role in determining the structure of the nucleus,
as was recognized and studied
in the earliest days of the nuclear shell model~\cite{Inglis53}.
The generally accepted conclusion is
that $jj$ coupling is relevant for the vast majority of nuclei
while the $LS$ classification
is appropriate for the very lightest nuclei only~\cite{Wilkinson95}.

Under certain circumstances a residual three-body interaction
must also be taken into account~\cite{Hammer13}.
In very light nuclei, the effects of three-body forces are especially important
and they must be incorporated fully~\cite{Navratil07}.
In heavier nuclei their effects are less pronounced
and can typically be absorbed into the mean field,
analogous to the two-body monopole interaction~\cite{Otsuka10}.

A second important feature
that plays a critical role in dictating the structure of the nucleus
is the number of active neutrons and protons
in the valence shell(s).
As noted earlier, the residual interaction between identical nucleons
has a pairing character which favours the formation
of pairs of nucleons in time-reversed orbits.
This is no longer true
if the valence space contains both neutrons and protons,
in which case there is a strong attraction in {\em all} orbits
resulting primarily through the quadrupole interaction [$J=2$ in equation~(\ref{e_vres})]
but with contributions under appropriate circumstances from other multipoles.
Hence, nuclei display a wide range of possible spectra
that can vary from pairing-type to rotational-like.
The evolution from one type to the other
is governed by the product of the number of neutrons
with the number of protons in the valence shell~\cite{Casten85}.

In heavier nuclei, {\it i.e.} in the rare-earth and actinide regions,
the mixing between orbitals with different parities becomes increasingly more important.
In these regimes important octupole-like ($J=3$) correlations arise,
and it is typically necessary to incorporate them
along with pairing and quadrupole correlations.

\subsection{A collective-model primer}
\label{ss_bm}
In 1879, in a study of the properties of a droplet of incompressible liquid,
Lord Rayleigh showed~\cite{Rayleigh79} that its normal modes of vibration
may be described by the variables $\alpha_{\lambda\mu}$
which appear in the expansion of the droplet's radius,
\begin{equation}
R(\theta,\phi)=R_0\left(1+\sum_{\lambda\mu}\alpha_{\lambda\mu}Y_{\lambda\mu}^*(\theta,\phi)\right),
\label{e_surface}
\end{equation}
where $Y_{\lambda\mu}(\theta,\phi)$ are spherical harmonics
in terms of the spherical angles $\theta$ and $\phi$.
In spite of some obvious differences
between a quantized atomic nucleus and a classical liquid drop,
the latter has been used to describe the properties of the atomic nucleus
since the pioneering work of von Weizs\"acker~\cite{Weizsacker35}
and Bohr and Kalckar~\cite{Bohr37}.
Since then it has been customary for nuclear physicists
to adopt the multipole parameterization~(\ref{e_surface})
to describe vibrations of the nuclear fluid.
There were several key steps and observations
that followed soon after these original works.
Feenberg~\cite{Feenberg39} and Bohr and Wheeler~\cite{Bohr39}
studied the shape and stability of a deformed and of a charged liquid drop.
A few years later Fl\"ugge~\cite{Flugge41} noted
that nuclear rotations may produce rotational spectra,
though he did not yet appreciate that surface vibrations and rotations
may be related to one another.
These ideas eventually culminated in the geometric collective model,
introduced in the classical papers
by Rainwater~\cite{Rainwater50}, Bohr~\cite{Bohr52},
and Bohr and Mottelson~\cite{Bohr53}.

The geometric collective model complements the spherical shell model
by emphasizing the coherent behaviour of many nucleons,
including quadrupole and higher-multipole deformations
as well as rotations and vibrations
that involve a large portion of the nucleus~\cite{Eisenberg70,Bohr75}.

\begin{figure}
\centering
\includegraphics[width=2.75cm]{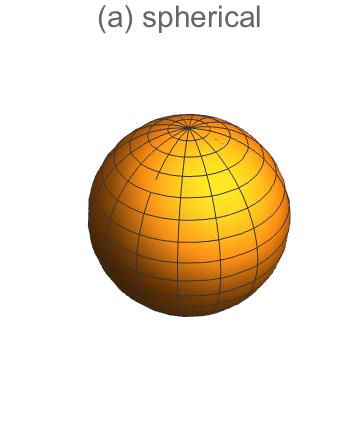}
\includegraphics[width=2.75cm]{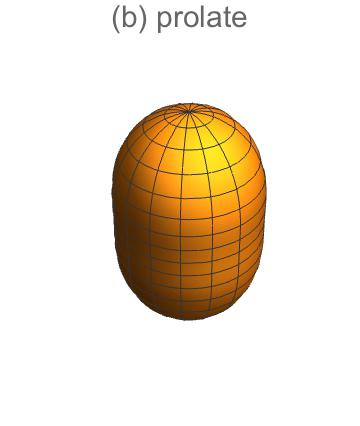}
\includegraphics[width=2.75cm]{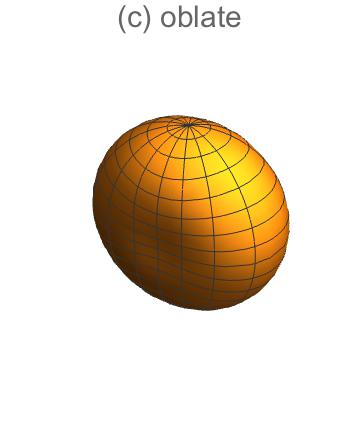}
\includegraphics[width=2.75cm]{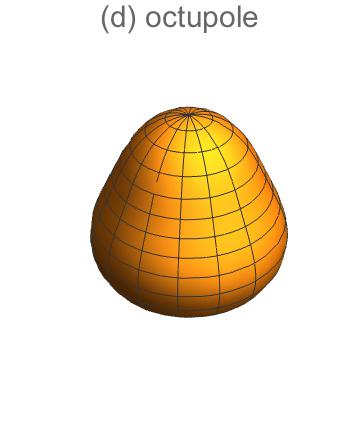}
\includegraphics[width=2.75cm]{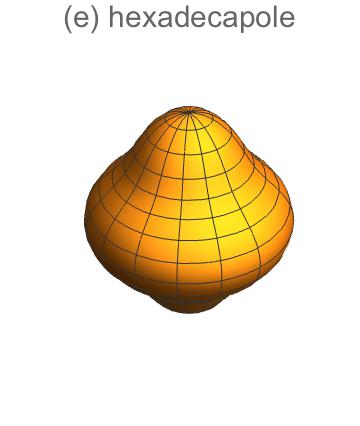}
\includegraphics[width=2.75cm]{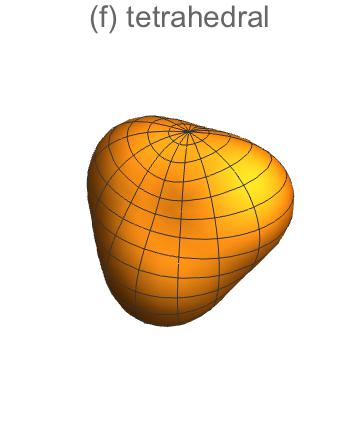}
\caption{Surfaces and their dependence on the variables $\alpha_{\lambda\mu}$.
The cases shown are
(a) spherical (all $\alpha_{\lambda\mu}$ are zero);
(b) prolate ($\alpha_{20}>0$);
(c) oblate ($\alpha_{20}<0$);
(d) octupole ($\alpha_{30}\neq0$)
(e) hexadecapole ($\alpha_{40}\neq0$)
and (f) tetrahedral (octupole with $\alpha_{32}\neq0$).}
\label{f_deformations}
\end{figure}

A key contribution of Bohr~\cite{Bohr52} and Bohr and Mottelson~\cite{Bohr53}
was to note that the static shape and orientation of a deformed nucleus
and the collective deformation variables of a spherical nucleus were related to each other.
Fl\"ugge had already suggested~\cite{Flugge41}
that these collective variables could play the role of dynamical variables
but it was Bohr~\cite{Bohr52} who extended this from a classical to a quantum drop.
In this picture the $\alpha_{\lambda\mu}$ in the nuclear surface~(\ref{e_surface})
are considered as (time-dependent) variables
that determine the shape of the nuclear surface.
For particular choices of $\lambda$ different shapes result.
This is illustrated in figure~\ref{f_deformations},
where the quadrupole case ($\lambda=2$) is shown
as well as examples of octupole ($\lambda=3$) 
and hexadecapole ($\lambda=4$) deformation.

For quadrupole deformations ($\lambda=2$),
which dominate in most regions of the periodic table,
the hamiltonian can be written as
\begin{equation}
\hat H=\hat T+\hat V=
\frac{1}{2B}\sum_\mu\left(\pi_{2\mu}\right)^2
+\frac{1}{2}C\sum_\mu\left(\alpha_{2\mu}\right)^2,
\label{e_hambm}
\end{equation}
where $\pi_{2\mu}$ is the momentum variable
associated with $\alpha_{2\mu}$, $\pi_{2\mu}=B\dot{\alpha}_{2\mu}$,
with $B$ the mass parameter and $C$ the restoring force.
The hamiltonian~(\ref{e_hambm})
corresponds to a five-dimensional harmonic oscillator
in the collective variables $\alpha_{2\mu}$
with frequency $\omega=\sqrt{C/B}$
and vibrational energy $\hbar\omega$.
After the introduction of the intrinsic coordinates $(\beta,\gamma)$
and the Euler angles,
its quantization leads to the well-known Bohr--Mottelson hamiltonian,
\begin{align*}
\hat H={}&-\frac{\hbar^2}{2B}\Bigg[
\frac{1}{\beta^4}\frac{\partial}{\partial\beta}
\beta^4\frac{\partial}{\partial\beta}+
\frac{1}{\beta^2}\Bigg(\frac{1}{\sin3\gamma}
\frac{\partial}{\partial\gamma}\sin3\gamma
\frac{\partial}{\partial\gamma}
\\&-
\frac{1}{4}\sum_\kappa
\frac{\hat I^{\prime2}_\kappa}{\sin^2(\gamma-2\pi\kappa/3)}\Bigg)
\Bigg]+\beta^2,
\end{align*}
where $\hat I'_\kappa$ are the components
of the angular momentum operator in the intrinsic frame of reference.
The solutions of this equation
are known in complete detail~\cite{Chacon76,Chacon77}.

While quadrupole deformation in its various manifestations
is prevalent throughout most of the periodic table,
some evidence for octupole deformation
can be found in the rare-earth and actinide regions.
While this evidence usually exists in terms of vibrational oscillations,
there are recent indications that the ground state of $^{224}$Ra
has a permanent octupole deformation~\cite{Gaffney13}.

The geometric collective model and its extensions
have been successful in describing a wide variety of nuclear properties.
Commonly measured properties,
including masses, angular momenta, magnetic moments and nuclear shapes,
can be understood from the geometric collective model.
Broad systematics of excited-state properties can likewise be described
but with important input from microscopic considerations required.

\section{Preamble: Wigner's supermultiplet model}
\label{s_wigner}
In Wigner's supermultiplet model~\cite{Wigner37}
nuclear forces are assumed to be invariant under rotations
in spin as well as isospin space.
A shell-model hamiltonian $\hat H$ with this property
satisfies the commutation relations
\begin{equation*}
[\hat H,\hat S_\mu]=
[\hat H,\hat T_\mu]=
[\hat H,\hat Y_{\mu\nu}]=0,
\end{equation*}
where
\begin{equation*}
\hat S_\mu=\sum_k\hat s_{k,\mu},
\quad
\hat T_\nu=\sum_k\hat t_{k,\nu},
\quad
\hat Y_{\mu\nu}=\sum_k\hat s_{k,\mu}\hat t_{k,\nu},
\end{equation*}
are the spin, isospin and spin--isospin operators
in terms of $\hat s_{k,\mu}$ and $\hat t_{k,\nu}$,
the spin and isospin components of nucleon $k$.
The set $\{\hat S_\mu,\hat T_\nu,\hat Y_{\mu\nu},\,\mu,\nu=1,2,3\}$
generates the Lie algebra SU(4)
and any hamiltonian that commutes with these 15 operators has SU(4) symmetry,
in addition to the SU(2) symmetries associated
with the total spin $S$ and total isospin $T$.

The physical relevance of Wigner's supermultiplet model
follows from the short-range attractive nature of the nuclear interaction,
which lowers the energy of states with increasing spatial symmetry.
This principle can be given a precise mathematical formulation
with the claim that the $n$-nucleon eigenstates
of a nuclear hamiltonian with SU(4) symmetry
are classified according to
\begin{equation}
\begin{array}{ccccc}
{\rm U}(4\Lambda)&\!\!\!\!\supset\!\!\!\!&{\rm U}_L(\Lambda)&\!\!\!\!\otimes\!\!\!\!&{\rm U}_{ST}(4)\\
\downarrow&&\downarrow&&\downarrow\\[0mm]
[1^n]&&[\bar h]&&[\bar h']
\end{array},
\label{e_su4a}
\end{equation}
where $\Lambda$ is the orbital dimension of the (valence) single-particle space,
$\Lambda=\sum_\ell(2\ell+1)$,
and $4\Lambda$ is the total single-particle dimension for neutrons and protons,
to account for the spin--isospin degrees of freedom.
The labels underneath the algebras are explained below.

The order of the algebras ${\rm U}_L(4\Lambda)$ and ${\rm U}_L(\Lambda)$
({\it i.e.}, their number of generators $16\Lambda^2$ and $\Lambda^2$)
is determined by the orbital shells $\ell$
that are included in the model space.
It is not necessary to specify what orbital shells are considered
but we assume in this section that summations over $\ell$
consistently include all of them.

For the subsequent discussion
it is convenient to use the formalism of second quantization 
and to introduce the operators $a^\dag_{\ell m_\ell sm_stm_t}$
which create a nucleon in the orbital shell $\ell$ with $z$ projection $m_\ell$,
spin $s={\frac12}$ with $z$ projection $m_s$
and isospin $t={\frac12}$ with $z$ projection $m_t$.
The corresponding annihilation operators are $a_{\ell m_\ell sm_stm_t}$
and, to ensure the correct transformation properties
under rotations in orbital, spin and isospin space,
one also introduces the modified annihilation operators
$\tilde a_{\ell m_\ell sm_stm_t}\equiv(-)^{\ell+m_\ell+s+m_s+t+m_t}$ $a_{\ell-m_\ell,s-m_s,t-m_t}$.
Anti-symmetry of the wave function is imposed
and the Pauli principle is respected
by requiring the following anti-commutation rules
among the fermion creation and annihilation operators:
\begin{align}
\{a^{\phantom\dag}_{\ell m_\ell sm_stm_t},a^\dag_{\ell'm'_\ell sm'_stm'_t}\}&=
\delta_{\ell\ell'}\delta_{m_\ell m'_\ell}\delta_{m_sm'_s}\delta_{m_tm'_t},
\nonumber\\
\{a_{\ell m_\ell sm_stm_t},a_{\ell'm'_\ell sm'_stm'_t}\}&=0,
\nonumber\\
\{a^\dag_{\ell m_\ell sm_stm_t},a^\dag_{\ell'm'_\ell sm'_stm'_t}\}&=0.
\label{e_anticom}
\end{align}

The generators of the algebra ${\rm U}(4\Lambda)$ in equation~(\ref{e_su4a})
can be written in terms of the coupled tensors
\begin{equation*}
(a^\dag_{\ell st}\times\tilde a^{\phantom\dag}_{\ell'st})^{(LST)}_{M_LM_SM_T},
\end{equation*}
where the superscripts
denote the coupling in orbital angular momentum $L$, spin $S$ and isospin $T$, respectively,
and the subscripts refer to their respective projections $M_L$, $M_S$ and $M_T$.
The explicit expression for the coupled tensors
involves ${\rm SO}(3)\supset{\rm SO}(2)$ coupling
or standard Clebsch--Gordan coefficients~\cite{Talmi93},
\begin{align*}
{}&(a^\dag_{\ell st}\times\tilde a^{\phantom\dag}_{\ell'st})^{(LST)}_{M_LM_SM_T}
\nonumber\\={}&
\sum_{m_\ell m'_\ell}(\ell m_\ell\,\ell'm'_\ell|LM_L)
\sum_{m_s m'_s}(s m_s\,sm'_s|SM_S)
\\&\times
\sum_{m_t m'_t}(t m_t\,tm'_t|TM_T)
a^\dag_{\ell m_\ell sm_stm_t}
\tilde a_{\ell'm'_\ell sm'_stm'_t}.
\end{align*}
It is assumed that all physical operators
({\it e.g.}, the hamiltonian, electromagnetic transition operators,\dots)
can be written in terms of the generators of ${\rm U}_L(4\Lambda)$.
Algebras with this property are
sometimes referred to as the dynamical
or spectrum generating algebra of the system under study~\cite{Bohm88,Frank09}.

The $\ell s$-coupled representation of nucleon creation and annihilation operators
is the most convenient for the purposes of the present paper.
Because of spin--orbit terms in the nucleon--nucleon interaction
it is no longer used in present-day shell-model studies
and is commonly replaced by a $jj$-coupled representation.
Both representations are nevertheless equivalent as a result of the relation
\begin{equation*}
a^\dag_{jm_jtm_t}=
\sum_{m_\ell m_s}
(\ell m_\ell\,sm_s|jm_j)
a^\dag_{\ell m_\ell sm_stm_t},
\end{equation*}
where $a^\dag_{jm_jtm_t}$ creates a nucleon 
in the shell $j$ with $z$ projection $m_j$
and with isospin $t={\frac12}$ with $z$ projection $m_t$.
The corresponding annihilation operators are $a_{jm_jtm_t}$
and the modified annihilation operators are defined as
$\tilde a_{jm_jtm_t}\equiv(-)^{j+m_j+t+m_t}$ $a_{j-m_j,t-m_t}$.

The generators of ${\rm U}_L(4\Lambda)$ can also
be written in terms of the $jj$-coupled tensors
\begin{equation*}
(a^\dag_{jt}\times\tilde a^{\phantom\dag}_{j't})^{(JT)}_{M_JM_T},
\end{equation*}
where the superscripts
denote the coupling in total ({\it i.e.}, $L$ plus $S$) angular momentum $J$ and isospin $T$, respectively,
and the subscripts refer to their respective projections $M_J$ and $M_T$.
Whether one uses the $\ell s$- or $jj$-coupled representation
is a matter of convenience
since both sets of generators are related by the unitary transformation
\begin{align*}
&(a^\dag_{\ell st}\times\tilde a^{\phantom\dag}_{\ell'st})^{(LST)}_{M_LM_SM_T}=
\sum_{JM_J}(LM_L\,SM_S|JM_J)
\nonumber\\&\qquad\qquad\qquad\times
\sum_{jj'}
\left[\!\!\begin{array}{ccc}
\ell&\frac12&j\\[0.5ex]
\ell'&\frac12&j'\\[0.5ex]
L&S&J
\end{array}\!\!\right]
(a^\dag_{jt}\times\tilde a^{\phantom\dag}_{j't})^{(JT)}_{M_JM_T},
\end{align*}
in terms of the unitary nine-$j$ symbol
\begin{equation*}
\left[\!\!\begin{array}{ccc}
j_1&j_2&J_{12}\\
j_3&j_4&J_{34}\\
J_{13}&J_{24}&J
\end{array}\!\!\right]\equiv
\hat J_{12}\hat J_{34}\hat J_{13}\hat J_{24}
\left\{\!\!\begin{array}{ccc}
j_1&j_2&J_{12}\\
j_3&j_4&J_{34}\\
J_{13}&J_{24}&J
\end{array}\!\!\right\},
\end{equation*}
where $\hat x\equiv\sqrt{2x+1}$
and the symbol in curly brackets is a standard nine-$j$ symbol~\cite{Talmi93}.

Two subalgebras of ${\rm U}_L(4\Lambda)$ appear in equation~(\ref{e_su4a}).
The first is ${\rm U}_L(\Lambda)$, which has the generators
\begin{equation}
\hat G^{(\lambda)}_\mu(\ell\ell')\equiv
(a^\dag_{\ell st}\times\tilde a^{\phantom\dag}_{\ell'st})^{(\lambda00)}_{\mu00},
\label{e_tensor}
\end{equation}
that is, coupled tensors that are scalar in spin and isospin,
as indicated by the superscripts $S=T=0$.
The angular momentum $\lambda$ runs
over all possible couplings of $\ell$ and $\ell'$
(which include all orbital shells of the model space),
$\lambda=|\ell-\ell'|,|\ell-\ell'|+1,\dots,\ell+\ell'$,
and $\mu$ is its projection,
$\mu=-\lambda,-\lambda+1,\dots,+\lambda$.
The generators of ${\rm U}_L(\Lambda)$
can also be written in terms of the operators
\begin{equation}
\hat G^{(\lambda)}_\mu(jj')\equiv
(a^\dag_{jt}\times\tilde a^{\phantom\dag}_{j't})^{(\lambda0)}_{\mu0}.
\label{e_tensorj}
\end{equation}
These  should not be confused with those in equation~(\ref{e_tensor})
since $j$ and $j'$ are always half-odd-integer,
in contrast to the integer values of $\ell$ and $\ell'$.
The operators~(\ref{e_tensorj}) generate the algebra ${\rm U}_L(2\Lambda)$,
which does not appear in the classification~(\ref{e_su4a})
but which contains the generators of ${\rm U}_L(\Lambda)$ as a subset since
\begin{equation*}
\hat G^{(\lambda)}_\mu(\ell\ell')=
\frac{(-)^{\ell+j'+1/2+\lambda}}{\sqrt2}\hat j \hat j'
\left\{\!\!\begin{array}{ccc}
j&\ell&\frac12\\\ell'&j'&\lambda
\end{array}\!\!\right\}
\hat G^{(\lambda)}_\mu(jj'),
\end{equation*}
where the symbol in curly brackets is a six-$j$ symbol~\cite{Talmi93}.
This relation allows us to write the $\ell s$-coupled tensors $\hat G^{(\lambda)}_\mu(\ell\ell')$ 
in terms of the $jj$-coupled tensors $\hat G^{(\lambda)}_\mu(jj')$,
which is more convenient for present-day applications.

The second subalgebra of ${\rm U}_L(4\Lambda)$ in equation~(\ref{e_su4a})
is ${\rm U}_{ST}(4)$ with the generators
\begin{equation*}
\sum_\ell(a^\dag_{\ell st}\times\tilde a^{\phantom\dag}_{\ell st})^{(0ST)}_{0M_SM_T},
\end{equation*}
which are scalar in orbital space
and with spin $S$ and isospin $T$ equal to 0 or 1.
The explicit definition of the ${\rm U}_{ST}(4)$ generators is 
\begin{align}
\hat n&\equiv
2\sum_\ell\sqrt{2\ell+1}(a^\dag_{\ell st}\times\tilde a^{\phantom\dag}_{\ell st})^{(000)}_{000},
\nonumber\\
\hat S_\mu&\equiv
\sum_\ell\sqrt{2(2\ell+1)}(a^\dag_{\ell st}\times\tilde a^{\phantom\dag}_{\ell st})^{(010)}_{0\mu0},
\nonumber\\
\hat T_\nu&\equiv
\sum_\ell\sqrt{2(2\ell+1)}(a^\dag_{\ell st}\times\tilde a^{\phantom\dag}_{\ell st})^{(001)}_{00\nu},
\nonumber\\
\hat Y_{\mu\nu}&\equiv
\sum_\ell\sqrt{2\ell+1}(a^\dag_{\ell st}\times\tilde a^{\phantom\dag}_{\ell st})^{(011)}_{0\mu\nu},
\label{e_genu4}
\end{align}
corresponding to the number, spin, isospin and spin--isospin operators
written in second quantization.

It can be shown
that the generators of ${\rm U}_L(\Lambda)$ and those of ${\rm U}_{ST}(4)$
close under commutation,
and that they commute which each other.
In particular, the following commutator property
among the generators of ${\rm U}_L(\Lambda)$ is valid:
\begin{align}
&[\hat G^{(\lambda)}_\mu(\ell_1\ell_2),\hat G^{(\lambda')}_{\mu'}(\ell_3\ell_4)]=
{\frac12}\hat\lambda\hat\lambda'\sum_{\lambda''\mu''}(\lambda\mu\,\lambda'\mu'|\lambda''\mu'')
\nonumber\\&\times\left[(-)^{\lambda''+\ell_1+\ell_4}
\left\{\!\!\begin{array}{ccc}
\lambda&\lambda'&\lambda''\\\ell_4&\ell_1&\ell_2
\end{array}\!\!\right\}
\delta_{\ell_2\ell_3}\hat G^{(\lambda'')}_{\mu''}(\ell_1\ell_4)\right.
\nonumber\\&\left.-(-)^{\lambda+\lambda'+\ell_2+\ell_3}
\left\{\!\!\begin{array}{ccc}
\lambda&\lambda'&\lambda''\\\ell_3&\ell_2&\ell_1
\end{array}\!\!\right\}
\delta_{\ell_1\ell_4}\hat G^{(\lambda'')}_{\mu''}(\ell_3\ell_2)\right].
\label{e_comula}
\end{align}
The (tedious) derivation of equation~(\ref{e_comula})
requires the expansion of the coupled tensors~(\ref{e_tensor})
in terms of uncoupled generators.
It makes use of the anti-commutators~(\ref{e_anticom})
and of summation properties of Clebsch--Gordan coefficients and six-$j$ symbols~\cite{Talmi93}.
The relation~(\ref{e_comula}) is identical to the corresponding one for bosons~\cite{Iachello06}
but for the factor 1/2 which originates from the coupling in spin and isospin.
The relation is central to the subsequent discussion
since many properties concerning various orbital classifications
can be derived from it.

Algebraic models in nuclear physics---and generally in quantum physics---rely
extensively on the notion of irreducible representation,
which is used to label eigenstates of hamiltonians with certain symmetries.
Because a system of $n$ identical particles
is invariant under permutations
that exchange all coordinates of any two particles,
representations of the permutation group ${\rm S}_n$,
consisting of all permutations of $n$ objects,
are of central importance.
As the system's hamiltonian is invariant under ${\rm S}_n$,
its eigenstates are labelled by the irreducible representations of ${\rm S}_n$.
If no condition is imposed
other than invariance under the exchange of {\em all} coordinates,
nothing more can be learned from permutation symmetry
than the fact that any nuclear eigenstate must be anti-symmetric.
The classification~(\ref{e_su4a}), however,
imposes the invariance under the exchange
of {\em only} the spatial (or, equivalently, {\em only} the spin--isospin)
coordinates of any two particles.
The symmetry character under such {\em partial} permutations
can be exploited to yield additional quantum numbers.

The symmetry type under a (total or partial) permutation
of a system of $n$ particles that occupy $\Lambda$ single-particle states
is characterized by a set of integers 
that satisfy the conditions $h_1\geq h_2\geq\cdots\geq h_\Lambda\geq0$
and $h_1+h_2+\cdots+h_\Lambda=n$.
This set of integers is denoted here as $[\bar h]\equiv[h_1,\dots,h_\Lambda]$
and is often represented as a Young pattern or diagram,
which corresponds to $n$ boxes
that are placed in $\Lambda$ rows of length $h_1,h_2,\dots$,
one underneath the other, beginning with $h_1$.
An irreducible representation of ${\rm S}_n$
characterized by a given Young diagram
contains basis states that are obtained
by placing each of the $n$ particles in a box
according to the following rule.
All particles are given a label between 1 and $n$.
They are then distributed over the boxes
such that in each row the particle index increases from left to right
and in each column the index increases from top to bottom.
The basis states obtained in this way are called Young tableaux.
For a given Young diagram each Young tableau 
corresponds to a different state with a given mixed symmetry,
which is obtained by anti-symmetrization
in the particles belonging to the same column,
after symmetrization in the particles
belonging to the same row (or {\it vice versa}).
In the case of complete anti-symmetry
the Young diagram reduces to a single column of $n$ boxes
($n\leq\Lambda$ because of the Pauli principle)
with only a single associated Young tableau,
namely the one with increasing particle index from top to bottom.
This shows that in the case of overall anti-symmetry
all states have an identical permutational character.

The mathematical theory of group representations 
is developed in the monographs
by Murnaghan~\cite{Murnaghan38} and Littlewood~\cite{Littlewood40}
while its applications to physical problems
are described in the treatise of Hamermesh~\cite{Hamermesh62}.
A clear and succinct account of the use of Young diagrams
in many-body quantum physics is given by Lipas~\cite{Lipas93}.

The states~(\ref{e_su4a}) are characterized
by a certain symmetry $[\bar h]\equiv[h_1,\dots,h_\Lambda]$ in orbital space
and a concomitant spin--isospin symmetry $[\bar h']\equiv[h'_1,h'_2,h'_3,h'_4]$,
with $h_1+\cdots+h_\Lambda=h'_1+h'_2+h'_3+h'_4=n$,
the total number of nucleons in the valence space.
The symmetry type under the partial exchange of coordinates
({\it e.g.}, the exchange of only the spatial or only the spin--isospin coordinates)
is thus characterized by a Young diagram.
To ensure the overall anti-symmetry of the wave function,
which is indicated in equation~(\ref{e_su4a})
by the one-column Young diagram $[1^n]\equiv[1,1,\dots,1]$ of ${\rm U}_L(4\Lambda)$,
the two Young diagrams $[\bar h]$ and $[\bar h']$
must be conjugate, that is,
they are obtained from each other by interchanging rows and columns.
%\deleted{If, for a given Young diagram $[\bar h]$ of ${\rm U}_L(\Lambda)$,
%the ${\cal S}_k^{[\bar h]}\equiv\{h_i \parallel h_i\geq k\}$
%denote the four sets of all labels $h_i$ greater than or equal to $k=1,\dots,4$,
%then the labels of the conjugate Young diagram $[\bar h']$ of ${\rm U}_{ST}(4)$
%are the cardinal numbers ({\it i.e.}, the number of elements) of those sets,
%$h'_k=|{\cal S}_k^{[\bar h]}|$.}
The conjugate relation is most easily explained with a figure:
A general Young diagram
\begin{equation*}
\begin{array}{l}
\overbrace{\fbox{}\,\fbox{}\,\fbox{}\dots\fbox{}}^{\textstyle h_1}\\
\overbrace{\fbox{}\,\fbox{}\dots\fbox{}}^{\textstyle h_2}\\
\;\vdots\\
\overbrace{\fbox{}\dots\fbox{}}^{\textstyle h_\Lambda}
\end{array},
\end{equation*}
becomes after conjugation
\begin{equation*}
\begin{array}{llll}
h_1\left\{\begin{array}{c}\\[-3.5ex]\fbox{}\\[-1ex]\fbox{}\\[-1ex]\fbox{}\\[-0.75ex]\,\vdots\\[-1.25ex]\fbox{}\end{array}\right.&
\begin{array}{c}
h_2\left\{\begin{array}{c}\\[-3.75ex]\fbox{}\\[-1ex]\fbox{}\\[-0.75ex]\,\vdots\\[-1.25ex]\fbox{}\end{array}\right.\\[-1ex]
\phantom{.}\end{array}&
\cdots&
\begin{array}{c}
h_\Lambda\left\{\begin{array}{c}\\[-4ex]\fbox{}\\[-0.75ex]\,\vdots\\[-1.25ex]\fbox{}\end{array}\right.\\
\phantom{.}\end{array}
\end{array}.
\end{equation*}
Since the spin--isospin algebra ${\rm U}_{ST}(4)$ is characterized
by a Young diagram of at most four rows,
it follows that the Young diagram associated with the orbital algebra ${\rm U}_L(\Lambda)$
has no more than four columns
or, equivalently, has rows with at most four boxes,
$4\geq h_1\geq h_2\geq\cdots\geq h_\Lambda\geq0$.
This is the group-theoretical transcription
of the fact that, in an anti-symmetric many-body wave function,
the same orbital single-particle state
cannot be occupied by more than four particles
corresponding to the four different nucleonic intrinsic states
with spin and isospin up or down.

The number operator $\hat n$ is common to both ${\rm U}_L(\Lambda)$ and ${\rm U}_{ST}(4)$,
and therefore ${\rm U}_L(\Lambda)\otimes{\rm U}_{ST}(4)$
is in fact not a direct product as the algebras involved are not disjoint.
This can be easily remedied
by considering the direct product ${\rm U}_L(\Lambda)\otimes{\rm SU}_{ST}(4)$,
where the number operator
is dropped from ${\rm U}_{ST}(4)$ to give ${\rm SU}_{ST}(4)$.
Irreducible representations of the latter algebra
are characterized by three labels,
which can be chosen as $[h'_1-h'_4,h'_2-h'_4,h'_3-h'_4]$
or, alternatively, as $(\lambda',\mu',\nu')$ with
\begin{equation*}
\lambda'\equiv h'_1-h'_2,
\quad
\mu'\equiv h'_2-h'_3,
\quad
\nu'\equiv h'_3-h'_4,
\end{equation*}
which corresponds to the more conventional notation of the supermultiplet model.
We conclude therefore that the classification~(\ref{e_su4a})
can be replaced by an equivalent one which reads
\begin{equation}
\begin{array}{ccccc}
{\rm U}(4\Lambda)&\!\!\!\!\supset\!\!\!\!&{\rm U}_L(\Lambda)&\!\!\!\!\otimes\!\!\!\!&{\rm SU}_{ST}(4)\\
\downarrow&&\downarrow&&\downarrow\\[0mm]
[1^n]&&[\bar h]&&(\lambda',\mu',\nu')
\end{array},
\label{e_su4b}
\end{equation}
where it is understood that the labels $(\lambda',\mu',\nu')$
are derived from the ${\rm U}_{ST}(4)$ labels $[\bar h']$,
which correspond to a Young diagram that is conjugate to $[\bar h]$.
In the following we use the classification~(\ref{e_su4a}) or~(\ref{e_su4b}),
whichever is most convenient,
with the understanding that both are equivalent.

A `supermultiplet' consists of all states
contained in an irreducible representation $[\bar h]$
of the orbital algebra ${\rm U}_L(\Lambda)$
or, equivalently, in an irreducible representation $(\lambda',\mu',\nu')$
of the spin--isospin algebra ${\rm SU}_{ST}(4)$.
The central idea of Wigner's supermultiplet model
is that the nucleon--nucleon interaction strongly favours
states with maximal spatial symmetry
and that as a consequence different supermultiplets are well separated in energy.
Low-energy states in the spectrum of a given nucleus
have maximal spatial symmetry
and therefore belong to the so-called `favoured supermultiplet'.

The separation of supermultiplets
can be achieved by an interaction of the form
\begin{equation}
\hat C_2[{\rm U}_L(\Lambda)]=
4\sum_{\ell\ell'}\sum_\lambda(-)^{\ell+\ell'}
\hat G^{(\lambda)}(\ell\ell')\cdot\hat G^{(\lambda)}(\ell'\ell),
\label{e_c2ula}
\end{equation}
where the dot denotes a scalar product,
\begin{equation*}
\hat T^{(\lambda)}\cdot\hat T^{(\lambda)}\equiv
(-)^\lambda\sqrt{2\lambda+1}\;
(\hat T^{(\lambda)}\times\hat T^{(\lambda)})^{(0)}_0.
\end{equation*}
The operator $\hat C_2[{\rm U}_L(\Lambda)]$ commutes
with all generators of ${\rm U}_L(\Lambda)$,
\begin{equation*}
[\hat G^{(\lambda)}_\mu(\ell\ell'),\hat C_2[{\rm U}_L(\Lambda)]]=0,
\end{equation*}
and therefore can be associated with the quadratic Casimir operator of ${\rm U}_L(\Lambda)$,
as anticipated by the notation.
To prove this commutator property use is made of equation~(\ref{e_comula})
together with the operator identity
\begin{equation*}
[\hat A,\hat B\hat C]=\hat B[\hat A,\hat C]+[\hat A,\hat B]\hat C.
\end{equation*}
The proof is straightforward,
to the extent that it can be delivered,
for particular realizations of ${\rm U}_L(\Lambda)$,
in a symbolic language like {\tt Mathematica}~\cite{Isackerun1}.
The presence of the phase factor $(-)^{\ell+\ell'}$
in the expression~(\ref{e_c2ula}) should be noted;
it has no importance if orbital shells of a single oscillator shell are included
but matters in case of orbital shells with both parities.

All bases considered in this paper
have the reduction~(\ref{e_su4a}) or~(\ref{e_su4b}) as a starting point.
Since the operator $\hat C_2[{\rm U}_L(\Lambda)]$ commutes
with all generators of ${\rm U}_L(\Lambda)$,
its eigenvalues in such bases
are known from classical group theory
(see, for example, table~5.1 of Ref.~\cite{Iachello06}) to be
\begin{equation}
\sum_{i=1}^\Lambda h_i(h_i+\Lambda+1-2i),
\label{e_eigula}
\end{equation}
where $h_i$ are the labels associated with ${\rm U}_L(\Lambda)$.
Casimir operators are only determined up to a proportionality factor
and the coefficient `4' in equation~(\ref{e_c2ula}) is chosen such
that the expectation value of $\hat C_2[{\rm U}_L(\Lambda)]$
yields the eigenvalue~(\ref{e_eigula}).

As the representations $[\bar h]$ and $[\bar h']$ are conjugate,
an entirely equivalent interaction can be proposed
in terms of the quadratic Casimir operator
of the spin--isospin algebra ${\rm U}_{ST}(4)$.
The operator $\hat C_2[{\rm U}_{ST}(4)]$ is also diagonal
in any basis associated with equation~(\ref{e_su4a}) or~(\ref{e_su4b})
and its eigenvalues are
\begin{equation*}
h'_1(h'_1+3)+h'_2(h'_2+1)+h'_3(h'_3-1)+h'_4(h'_4-3),
\end{equation*}
where $h'_i$ are the labels associated with ${\rm U}_{ST}(4)$.
Given that $[\bar h]$ and $[\bar h']$ are conjugate Young diagrams,
the following relation between the two Casimir operators can be established:
\begin{equation*}
\hat C_2[{\rm U}_L(\Lambda)]=(\Lambda+4)\hat n-\hat C_2[{\rm U}_{ST}(4)].
\end{equation*}
We emphasize that this relation is {\em not} generally valid
but that it applies only in the anti-symmetric representation $[1^n]$ of ${\rm U}(4\Lambda)$
since it uses the fact that $[\bar h]$ and $[\bar h']$ are conjugate.

For completeness we also quote the expression
for the quadratic Casimir operator of ${\rm SU}_{ST}(4)$.
In terms of the labels $(\lambda',\mu',\nu')$
the eigenvalue of $4\hat C_2[{\rm U}_{ST}(4)]$ is rewritten as
\begin{equation*}
3\lambda'(\lambda'+4)+
4\mu'(\mu'+4)+
3\nu'(\nu'+4)+
4\mu'(\lambda'+\nu')+
2\lambda'\nu'+n^2.
\end{equation*}
Therefore, since $n$ is a constant for a given nucleus,
we may define
\begin{equation*}
\hat C_2[{\rm SU}_{ST}(4)]\equiv4\hat C_2[{\rm U}_{ST}(4)]-\hat n^2,
\end{equation*}
and we find that the Casimir operator $\hat C_2[{\rm SU}_{ST}(4)]$
has the eigenvalues
\begin{equation}
3\lambda'(\lambda'+4)+
4\mu'(\mu'+4)+
3\nu'(\nu'+4)+
4\mu'(\lambda'+\nu')+
2\lambda'\nu',
\label{e_eigsu4}
\end{equation}
which corresponds to the conventional expression of the supermultiplet model,
as quoted for example in Ref.~\cite{Elliott81}.

States with maximal spatial symmetry
have the largest possible eigenvalue of $\hat C_2[{\rm U}_L(\Lambda)]$
or, equivalently, the smallest possible eigenvalue
of $\hat C_2[{\rm U}_{ST}(4)]$ or $\hat C_2[{\rm SU}_{ST}(4)]$.
For a given nucleus
with $N$ neutrons and $Z$ protons,
and isospin projection $T_z=(N-Z)/2$,
the favoured supermultiplet must be compatible
with the minimal isospin $T=|T_z|$ of states in that nucleus.
The allowed values of the total spin $S$ and the total isospin $T$
in a given supermultiplet $(\lambda',\mu',\nu')$
are found from the branching rule associated with
\begin{equation}
\begin{array}{ccccc}
{\rm SU}_{ST}(4)&\!\!\!\!\supset\!\!\!\!&{\rm SU}_S(2)&\!\!\!\!\otimes\!\!\!\!&{\rm SU}_T(2)\\
\downarrow&&\downarrow&&\downarrow\\[0mm]
(\lambda',\mu',\nu')&&S&&T
\end{array}.
\label{e_redsu4}
\end{equation}
Most cases of interest for this branching rule have been tabulated
(see, for example, Refs.~\cite{Hecht69b,Elliott81}).
All cases can be found starting from the equivalent branching rule for
\begin{equation*}
\begin{array}{ccccc}
{\rm U}_{ST}(4)&\!\!\!\!\supset\!\!\!\!&{\rm U}_S(2)&\!\!\!\!\otimes\!\!\!\!&{\rm U}_T(2)\\
\downarrow&&\downarrow&&\downarrow\\[0mm]
[\bar h']&&[s_1,s_2]&&[t_1,t_2]
\end{array},
\end{equation*}
with the relation between $(\lambda',\mu',\nu')$ and $[\bar h']$
as explained above,
and with $S=(s_1-s_2)/2$ and $T=(t_1-t_2)/2$.
This is a particular case of the branching rule for
${\rm U}(n_1n_2)\supset{\rm U}(n_1)\otimes{\rm U}(n_2)$,
the algorithm of which is explained in the appendix.

The favoured supermultiplet for a given nucleus
is therefore found from the following procedure:
\begin{itemize}
\item
Determine the number $n$ of neutrons and protons in the valence shell.
\item
For that value of $n$, enumerate all possible Young diagrams $[\bar h]$
associated with ${\rm U}_L(\Lambda)$.
This amounts to finding all partitions of $n$ into $\Lambda$ integers
$h_1,h_2,\dots,h_\Lambda$ that satisfy
$4\geq h_1\geq h_2\geq\cdots\geq h_\Lambda\geq0$.
\item
For each $[\bar h]$ find the conjugate Young diagram $[\bar h']$,
the corresponding labels $(\lambda',\mu',\nu')$
and the allowed values of $S$ and $T$ from the branching rule~(\ref{e_redsu4}).
\item
The favoured supermultiplet corresponds
to the ${\rm SU}_{ST}(4)$ irreducible representation $(\lambda',\mu',\nu')$
with the smallest possible eigenvalue~(\ref{e_eigsu4})
which contains the isospin $T=|T_z|$.
\end{itemize}

\begin{table*}
\centering
\caption{Favoured  SU(4) supermultiplets of nuclei with $n$ valence nucleons.}
\label{t_favsu4}
\smallskip
\begin{tabular}{lll}
\hline\hline
Nucleus&$[\bar h']^a$&$(\lambda',\mu',\nu')$\\
\hline
Even--even&$[k+|T_z|,k+|T_z|,k,k]$&$(0,|T_z|,0)$\\
Odd-mass&$[k+|T_z|+{\frac12},k+|T_z|-{\frac12},k,k]$&$(1,|T_z|-{\frac12},0)$\\
Odd--odd $(N\neq Z)$&$[k'+|T_z|+1,k'+|T_z|,k'+1,k']$&$(1,|T_z|-1,1)$\\
Odd--odd $(N=Z)$&$[k'+1,k'+1,k',k']$&$(0,1,0)$\\
\hline\hline
\multicolumn{3}{l}{$^a$With $k=(n-2|T_z|)/4$; $k'=(n-2|T_z|-2)/4$.}
\end{tabular}
\end{table*}
Through the application of the above procedure,
generic rules emerge that determine the favoured supermultiplet for any nucleus,
depending on whether it is even--even, odd-mass or odd--odd.
From the summary shown in table~\ref{t_favsu4} it is seen
that the ${\rm U}_{ST}(4)$ labels $[\bar h']$
are functions of the number of valence nucleons $n$
and the isospin projection $|T_z|$
while the ${\rm SU}_{ST}(4)$ labels $(\lambda',\mu',\nu')$
more conveniently only depend on $|T_z|$.
Once $[\bar h']$ is determined,
the orbital symmetry $[\bar h]$ follows from conjugation.

We conclude the discussion of Wigner's supermultiplet model
with two comments concerning $N=Z$ nuclei.
The above procedure does not {\it a priori} exclude
that an isospin larger than $|T_z|$
is also contained in the favoured supermultiplet.
With one exception this in fact never happens.
In other words, states with $T>|T_z|$
almost always belong to the next-favoured supermultiplet
and, on account of this finding, are well separated in energy from states with $T=|T_z|$.
The one exception concerns odd--odd $N=Z$ nuclei:
According to table~\ref{t_favsu4} the favoured supermultiplet in that case
is $(\lambda',\mu',\nu')=(0,1,0)$,
which contains both $T=0$ {\em and} $T=1$.
Low-energy levels in odd--odd $N=Z$ nuclei, uniquely, indeed carry both isospins.
This empirical observation,
the explanation of which traditionally invokes convoluted arguments
about counterbalancing effects of pairing and symmetry energies~\cite{Vogel00,Frauendorf14},
can be explained naturally in the framework of the SU(4) model.

The second comment concerns the so-called Wigner binding energy.
Self-conjugate nuclei, that is,
nuclei with an equal number of neutrons and protons ($N=Z$),
are unusually tightly bound.
This extra binding energy was first noted by Wigner
who proposed an explanation of the observed ``kinks in the mass defect curve''
with symmetry arguments~\cite{Wigner37b}.
Since then, the $N=Z$ cusp in the nuclear mass surface
is often described with an additional term in binding-energy formulas,
known as the Wigner binding energy.

The Wigner binding energy $B_{\rm W}(N,Z)$
consists of two parts~\cite{Moller92}
\begin{equation*}
B_{\rm W}(N,Z)=-W(A)|N-Z|-d(A)\delta_{N,Z}\pi_{\rm np},
\end{equation*}
where $W(A)$ and $d(A)$ are functions of the atomic mass number $A\equiv N+Z$.
The first term on the right-hand side
gives rise to a cusp in the binding energy at $N=Z$.
By definition, the quantity $\pi_{\rm np}$
equals 1 for odd--odd nuclei and vanishes otherwise,
and therefore $d(A)$ matters only for odd--odd $N=Z$ nuclei.
Empirical estimates lead to $W(A)\approx47A^{-1}$~MeV
and $d(A)$ comparable but possibly somewhat smaller~\cite{Satula97}.

The origin of the two terms in the Wigner binding energy
can be explained on the basis of previously derived results.
Insertion of the favoured supermultiplet labels $(\lambda',\mu',\nu')$,
as given in table~\ref{t_favsu4},
into the eigenvalue~(\ref{e_eigsu4}) of $\hat C_2[{\rm SU}_{ST}(4)]$
leads to the expression~\cite{Warner06}
\begin{equation*}
(N-Z)^2+8|N-Z|+8\delta_{N,Z}\pi_{\rm np}+6\delta_{\rm p}(N,Z).
\end{equation*}
One recovers the classical symmetry energy
of nuclear mass formulas in the first term.
The last term has a pairing-like character
and follows the somewhat unusual convention
that $\delta_{\rm p}(N,Z)$ is 0 for even--even,
1 for odd-mass and 2 for odd--odd nuclei.
The second and third terms correspond exactly
to those in $B_{\rm W}(N,Z)$ with the constraint $W(A)=d(A)=8g(A)$,
where $g(A)$ is the coefficient in front of $\hat C_2[{\rm SU}_{ST}(4)]$.
(The change in sign follows from the fact
that the eigenvalue of $\hat C_2[{\rm SU}_{ST}(4)]$
refers to an {\em interaction} energy between nucleons
whereas $B_{\rm W}(N,Z)$ is a {\em binding} energy.)
We therefore conclude that the Wigner term $B_{\rm W}(N,Z)$ in nuclear mass formulas
is directly inspired by the supermultiplet model.

Wigner's supermultiplet model is based on an $LS$-coupling scheme
which is now known to be inappropriate for most nuclei.
The breakdown of SU(4) symmetry
is a consequence of the spin- and isospin-dependent (Coulomb) interactions,
in particular the spin--orbit term in the nuclear mean field.
Nevertheless, Wigner's idea remains relevant
because it illustrates the connection
between the short-range character of the nuclear interaction
and the spatial symmetry of the many-body wave function.
One of the most important ramifications of the supermultiplet model
is its extension to include rotational motion
in the spherical shell model by way of Elliott's SU(3) model.

\section{Elliott's SU(3) model of quadrupole deformation}
\label{s_elliott}
In this section we review Elliott's SU(3) model~\cite{Elliott58},
which can be considered as a further elaboration upon the supermultiplet scheme.
A convenient way to understand this symmetry
takes the isotropic harmonic oscillator as a starting point,
whose degeneracies can be interpreted in terms of a U(3) algebra~\cite{Jauch40}.
A natural realization of this algebra is in a cartesian basis
but, as shown by Elliott~\cite{Elliott58},
an equivalent representation is possible in a spherical basis
and leads to a realization of U(3)
in terms of a number, an angular momentum and a quadrupole operator.

The subsequent presentation of the SU(3) model
differs somewhat from that in Elliott's original papers
in the sense that we assume,
as is customary in present-day shell-model calculations,
a given single-particle valence space,
in which the many-body nuclear hamiltonian is diagonalized.
From this point of view,
Elliott's result can be succinctly summarized as follows.
If the single-particle space consists of one or several entire, degenerate oscillator shells
and if the interaction between the nucleons in that space
is exclusively of the quadrupole type,
then an SU(3) dynamical symmetry results.
We first present the analysis for one oscillator shell
and then discuss extensions to more shells.

\subsection{One oscillator shell}
\label{ss_quad1}
For nucleons occupying an entire shell of the harmonic oscillator in three dimensions,
with major quantum number $N$ and orbital shells $\ell=N,N-2,\dots,1$ or 0,
the following generic orbital classification can be proposed:
\begin{equation}
\begin{array}{ccccccc}
{\rm U}(\Gamma)&\!\!\!\!\supset\!\!\!\!&{\rm U}(3)&\!\!\!\!\supset\!\!\!\!&{\rm SU}(3)&\!\!\!\!\supset\!\!\!\!&{\rm SO}(3)\\
\downarrow&&\downarrow&&\downarrow&&\downarrow\\[0mm]
[\bar h]&&[\bar h'']&&(\lambda,\mu)&K&L
\end{array},
\label{e_clasq}
\end{equation}
where the subscript `$L$' is omitted from the orbital algebras for simplicity's sake
and the notation $\Gamma\equiv (N+1)(N+2)/2$ is introduced.
The algebra U(3) consists
of the number operator $\hat n$ [see equation~(\ref{e_genu4})],
the three components of the angular momentum operator,
\begin{equation}
\hat L_\mu\equiv\sum_\ell
\sqrt{\frac{4\ell(\ell+1)(2\ell+1)}{3}}
\hat G^{(1)}_\mu(\ell\ell),
\label{e_ang}
\end{equation}
and the five components of the quadrupole operator,
\begin{equation}
\hat Q_\mu\equiv\sqrt\frac{64\pi}{5}\sum_{\ell\ell'}
\frac{\langle\ell||r^2 Y_2||\ell'\rangle}{\sqrt5}
\hat G^{(2)}_\mu(\ell\ell'),
\label{e_quad}
\end{equation}
where it is assumed
that the summations are over $\ell,\ell'=N,N-2,\dots,1$ or 0.
The algebra SU(3) in the orbital classification~(\ref{e_clasq})
consists of $\hat L_\mu$ and $\hat Q_\mu$.
The angular momentum operator $\hat L_\mu$ has a fixed structure,
independent of the potential well,
while the expression for the quadrupole operator $\hat Q_\mu$
involves the reduced matrix elements $\langle\ell||r^2 Y_2||\ell'\rangle$
that do depend on radial integrals.
These are well known for a harmonic oscillator~\cite{Talmi93},
\begin{align*}
&\langle\ell||r^2 Y_2||\ell\rangle=
-(2N+3)\sqrt{\frac{5}{16\pi}\frac{\ell(\ell+1)(2\ell+1)}{(2\ell-1)(2\ell+3)}},
\\
&\langle\ell||r^2 Y_2||\ell+2\rangle
\\
&\qquad=-\sqrt{\frac{5}{16\pi}\frac{6(\ell+1)(\ell+2)(N-\ell)(N+\ell+3)}{2\ell+3}}.
\nonumber
\end{align*}
It was shown by Elliott~\cite{Elliott58}
that the set of eight operators $\{\hat L_\mu,\hat Q_\mu\}$,
pertaining to the harmonic oscillator,
close under commutation
and hence form a subalgebra of ${\rm U}(\Gamma)$.
The commutation relations
\begin{align}
[\hat L_\mu,\hat L_\nu]&=
-\sqrt{2}\,(1\mu\,1\nu|1\mu+\nu)\hat L_{\mu+\nu},
\nonumber\\[0mm]
[\hat L_\mu,\hat Q_\nu]&=
-\sqrt{6}\,(1\mu\,2\nu|2\mu+\nu)\hat Q_{\mu+\nu},
\nonumber\\[0mm]
[\hat Q_\mu,\hat Q_\nu]&=
3\sqrt{10}\,(2\mu\,2\nu|1\mu+\nu)\hat L_{\mu+\nu},
\label{e_comsu3}
\end{align}
follow from the straightforward application of the commutator property~(\ref{e_comula})
and can be proven, for particular realizations of ${\rm U}(\Gamma)$
({\it i.e.}, for particular values of $N$),
in a symbolic language like {\tt Mathematica}~\cite{Isackerun1}.

In the orbital classification~(\ref{e_clasq})
appear the ${\rm U}(\Gamma)$ labels $[\bar h]$
that are carried over from Wigner's supermultiplet model.
The algebra U(3) is characterized
by a three-rowed Young diagram $[\bar h'']\equiv[h''_1,h''_2,h''_3]$
while the conventional notation for the SU(3) labels is $(\lambda,\mu)$ with
\begin{equation*}
\lambda\equiv h''_1-h''_2,
\quad
\mu\equiv h''_2-h''_3.
\end{equation*}
The label $L$ in equation~(\ref{e_clasq})
refers to the total orbital angular momentum of the nucleons
and an additional index $K$ occurs
not associated with any algebra
but needed as a multiplicity label in the branching rule for ${\rm SU}(3)\supset{\rm SO}(3)$.
The allowed Young diagrams $[\bar h'']$ of ${\rm U}(3)$
or, equivalently, the allowed SU(3) labels $(\lambda,\mu)$
for a given Young diagram $[\bar h]$ of ${\rm U}(\Gamma)$
are determined from the branching rule for ${\rm U}(\Gamma)\supset{\rm U}(3)$,
for which a general method exists known as the plethysm of $S$ functions~\cite{Littlewood40}.
Furthermore, the branching rule for ${\rm SU}(3)\supset{\rm SO}(3)$
can be obtained from $S$-function theory
for restricted ({\it i.e.}, symplectic or orthogonal) algebras.
This determines the allowed values of $L$
in a given irreducible representation $(\lambda,\mu)$.

Group-theoretical methods related to $S$-function theory
are described, for example, in the book by Wybourne~\cite{Wybourne70}.
A succinct summary of these methods,
which suffices for the applications discussed in this paper,
is given in the appendix.

Rather than giving the actual branching rules,
which can be found in the original papers by Elliott~\cite{Elliott58},
we note that each irreducible representation $(\lambda,\mu)$
contains the orbital angular momenta $L$ typical of a rotational band,
cut off at some upper limit.
The label $K$ defines the intrinsic state associated to that band
and can be interpreted as the projection of the orbital angular momentum $L$
on the axis of symmetry of the rotating deformed nucleus.

\begin{figure*}
\centering
\includegraphics[height=6.5cm]{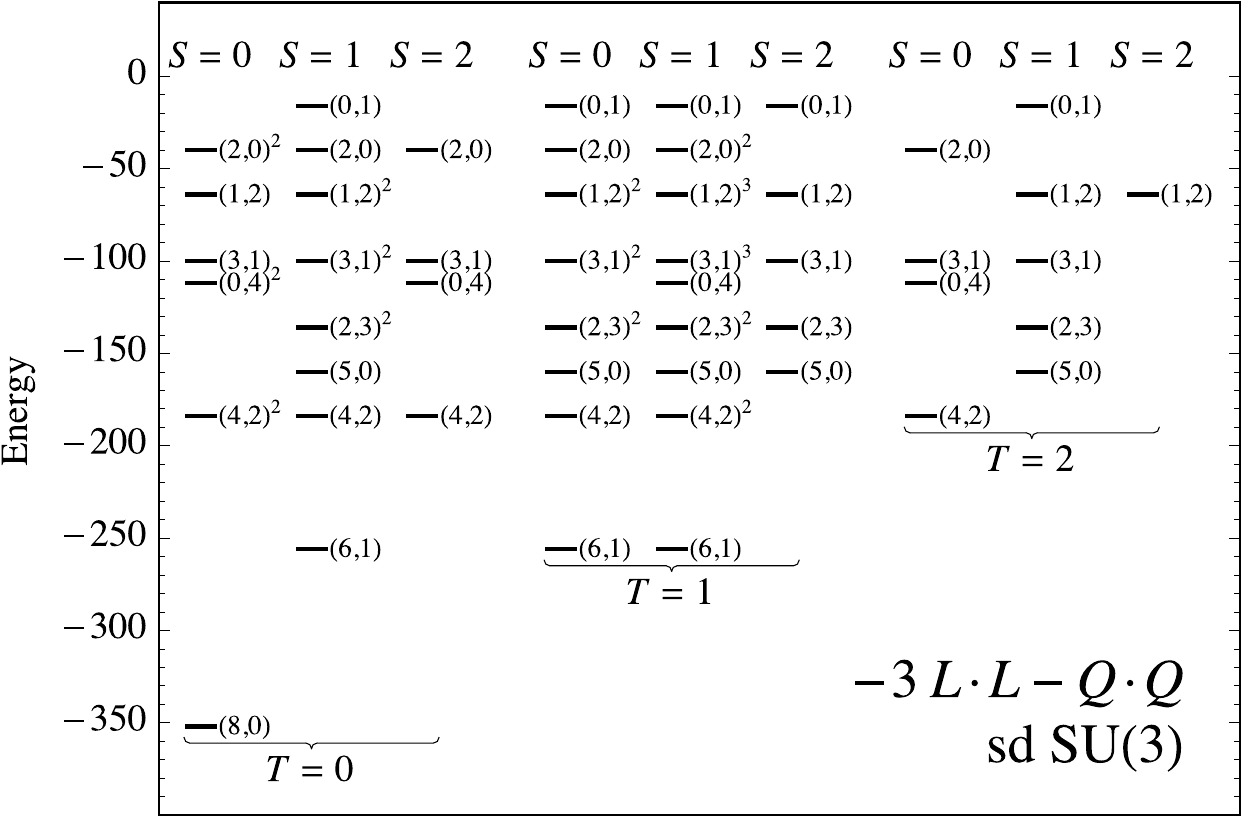}
\includegraphics[height=6.5cm]{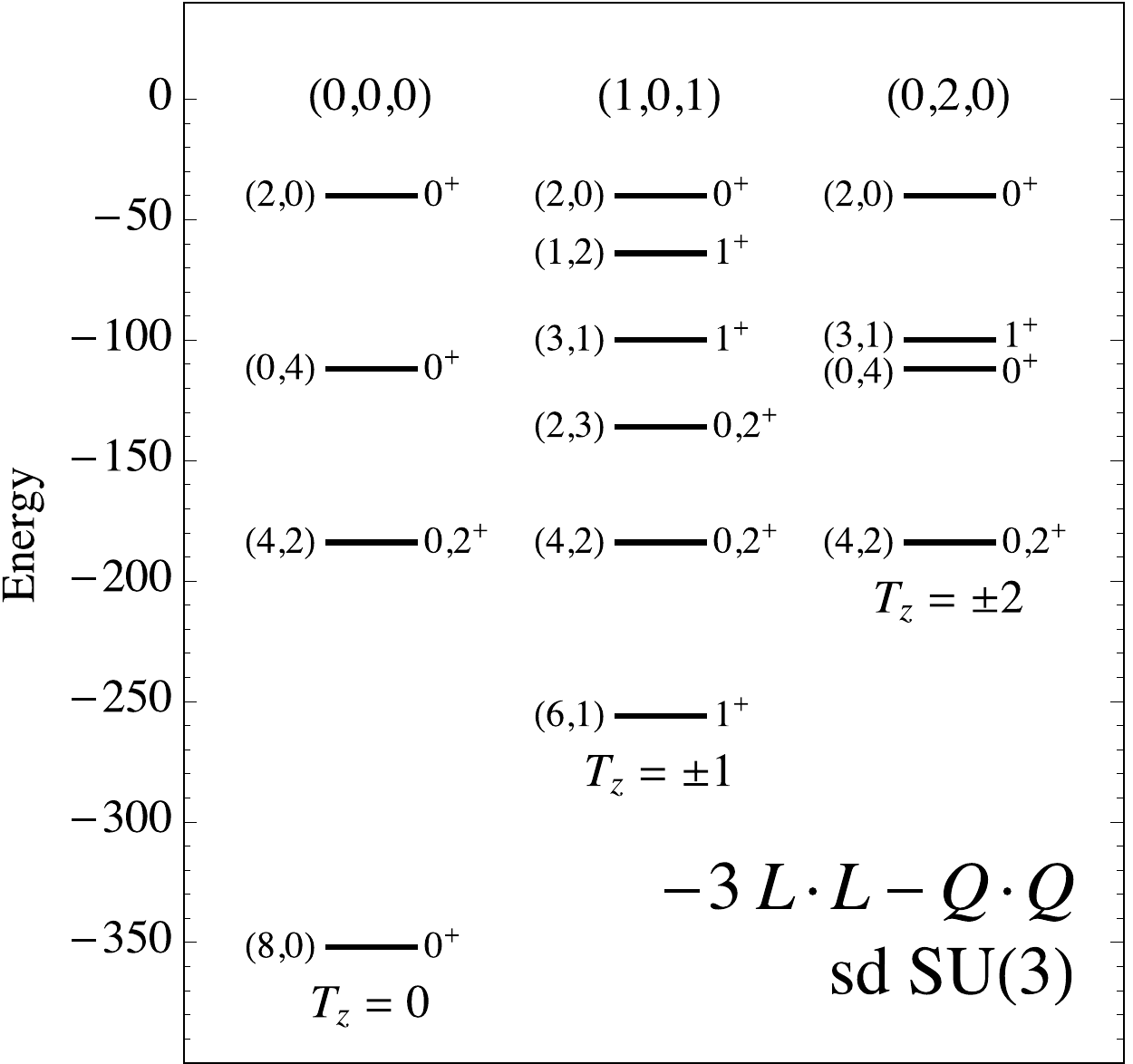}
\caption{The eigenspectrum of the operator $-3\hat L\cdot\hat L-\hat Q\cdot\hat Q$
for four nucleons in the $sd$ shell.
Left panel:
Levels are labelled by the SU(3) quantum numbers $(\lambda,\mu)$
and values of the total spin $S$ and the total isospin $T$ are also indicated.
If an irreducible representation $(\lambda,\mu)$
occurs $k$ times for a given $S$ and $T$,
this is indicated by a superscript as $(\lambda,\mu)^k$.
Right panel:
Only levels in the favoured supermultiplets are shown,
which are $(0,0,0)$ for $T_z=0$,
$(1,0,1)$ for $T_z=\pm1$
and $(0,2,0)$ for $T_z=\pm2$.
Levels are labelled
by the SU(3) quantum numbers $(\lambda,\mu)$ on the left
and by the projections $K$ and parity $\pi=+$ on the right.
The supermultiplet labels $(\lambda',\mu',\nu')$
and the isospin projection $T_z$ are also indicated,
and all levels have $S=0$.}
\label{f_quad1}
\end{figure*}
The combination $2\hat n^2+3\hat L\cdot\hat L+\hat Q\cdot\hat Q$
commutes with all generators of U(3)
and hence can be identified with the quadratic Casimir operator of U(3),
\begin{equation*}
\hat C_2[{\rm U}(3)]={\frac13}\hat n^2+{\frac12}\hat L\cdot\hat L+{\frac16}\hat Q\cdot\hat Q,
\end{equation*}
which according to equation~(\ref{e_eigula}) has the eigenvalues
\begin{equation*}
h''_1(h''_1+2)+h_2^{\prime\prime2}+h''_3(h''_3-2).
\end{equation*}
In terms of the labels $(\lambda,\mu)$ of SU(3)
the eigenvalue of $3\hat C_2[{\rm U}(3)]$ is rewritten as
\begin{equation*}
2\lambda(\lambda+3)+2\mu(\mu+3)+2\lambda\mu+n^2.
\end{equation*}
Therefore, since $n$ is a constant for a given nucleus,
we may define
\begin{equation}
\hat C_2[{\rm SU}(3)]=
3\hat C_2[{\rm U}(3)]-\hat n^2=
{\frac32}\hat L\cdot\hat L+{\frac12}\hat Q\cdot\hat Q,
\label{e_c2su3}
\end{equation}
which has the eigenvalues
\begin{equation}
2\lambda(\lambda+3)+2\mu(\mu+3)+2\lambda\mu.
\label{e_eigsu3}
\end{equation}
The quadrupole interaction is thus a combination of Casimir operators,
\begin{equation}
-\hat Q\cdot\hat Q=
-2\hat C_2[{\rm SU}(3)]+3\hat C_2[{\rm SO}(3)].
\label{e_hamq}
\end{equation}
Since this hamiltonian can be written
as a combination of Casimir operators
belonging to the chain~(\ref{e_clasq}) of nested algebras,
it is solvable with eigenstates
\begin{equation}
|[1^n];[\bar h](\lambda,\mu)KL\times[\bar h']ST\rangle,
\label{e_basq}
\end{equation}
and energy eigenvalues
\begin{equation}
-4\lambda(\lambda+3)-4\mu(\mu+3)-4\lambda\mu+3L(L+1).
\label{e_eigq}
\end{equation}

We illustrate the procedure
to obtain the complete eigenspectrum of the Casimir operator~(\ref{e_c2su3})
with the example of four nucleons in the $sd$ shell
({\it i.e.}, $n=4$ and $\Gamma=6$).
The allowed U(6) labels $[\bar h]$ correspond to all partitions of $n=4$
and they are thus given by
\begin{equation*}
[\bar h]=[4],[3,1],[2,2],[2,1,1],[1,1,1,1].
\end{equation*}
The equivalent series of U(4) labels
consists of the conjugate Young diagrams,
\begin{equation*}
[\bar h']=[1,1,1,1],[2,1,1],[2,2],[3,1],[4].
\end{equation*}
The former series determines the allowed SU(3) labels $(\lambda,\mu)$,
which follow from the branching rules for ${\rm U}(6)\supset{\rm SU}(3)$
and can be taken from Elliott~\cite{Elliott58}
(see also the appendix),
\begin{align*}
[4]\mapsto{}&(8,0)+(4,2)+(0,4)+(2,0),
\\
[3,1]\mapsto{}&(6,1)+(4,2)+(2,3)+(3,1)+(1,2)+(2,0),
\\
[2^2]\mapsto{}&(4,2)+(0,4)+(3,1)+(2,0),
\\
[2,1^2]\mapsto{}&(5,0)+(2,3)+(3,1)+(1,2)+(0,1),
\\
[1^4]\mapsto{}&(1,2),
\end{align*}
while the latter series determines the allowed $(S,T)$ values,
which follow from the branching rules for ${\rm U}(4)\supset{\rm SU}(2)\otimes{\rm SU}(2)$
(see Ref.~\cite{Hecht69b} and also the appendix),
\begin{align*}
[1^4]\mapsto{}&(0,0),
\\
[2,1^2]\mapsto{}&(0,1)+(1,0)+(1,1),
\\
[2^2]\mapsto{}&(0,0)+(1,1)+(0,2)+(2,0),
\\
[3,1]\mapsto{}&(0,1)+(1,0)+(1,1)+(1,2)+(2,1),
\\
[4]\mapsto{}&(0,0)+(1,1)+(2,2).
\end{align*}
By combining the information from both types of branching rule,
we know therefore for each of the nine possible $(S,T)$ combinations
what are the allowed SU(3) labels and how many times they occur.

\begin{figure}
\centering
\includegraphics[width=8cm]{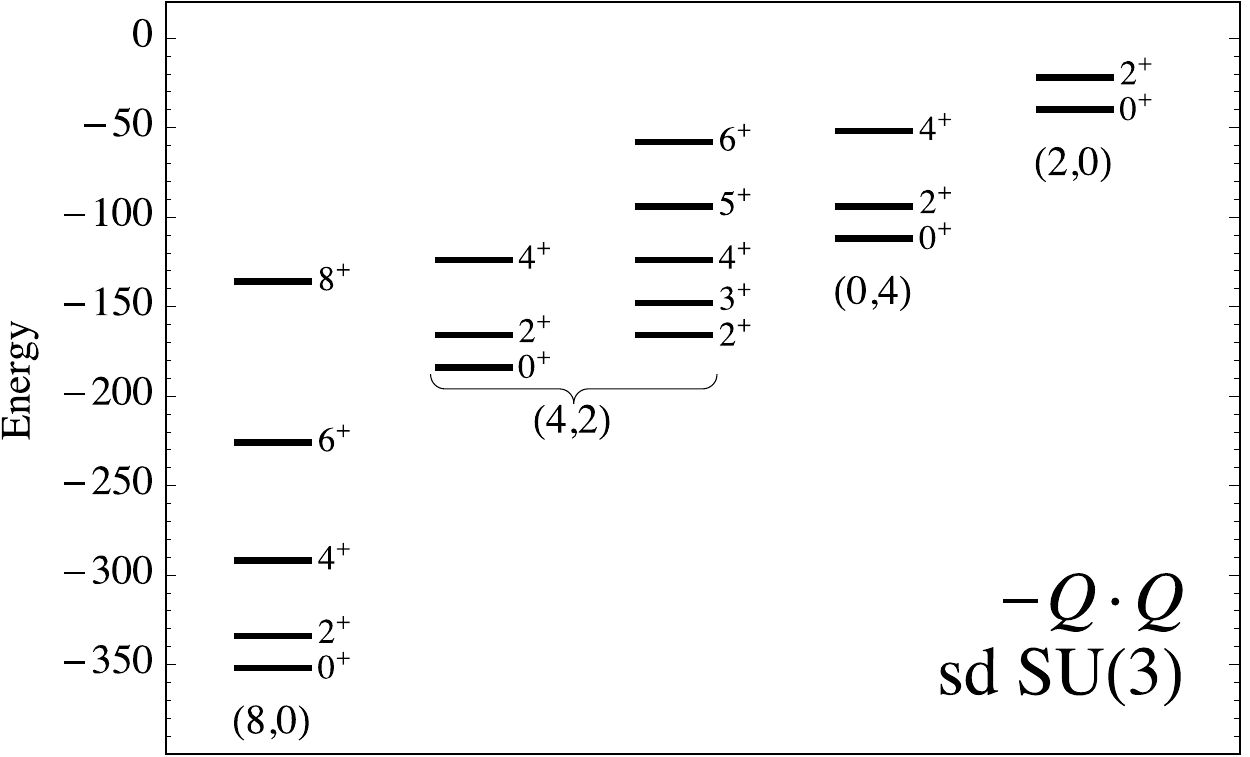}
\caption{The eigenspectrum of the operator $-\hat Q\cdot\hat Q$
for two neutrons and two protons in the $sd$ shell.
Only levels in the favoured supermultiplet $(0,0,0)$ are shown.
Levels are labelled by the orbital angular momentum $L$ and parity $\pi=+$,
and by the SU(3) quantum numbers $(\lambda,\mu)$.
All levels have $S=0$
and therefore the total angular momentum $J$ equals the orbital angular momentum $L$.}
\label{f_quad1c}
\end{figure}
The resulting eigenspectrum is displayed in figure~\ref{f_quad1}.
The spectrum in the left panel is complete
and shows all irreducible representations $(\lambda,\mu)$ with their multiplicities
in the five possible supermultiplets.
As argued in section~\ref{s_wigner},
the nuclear interaction lowers the energy of the states with maximal spatial symmetry
that occur in the favoured supermultiplet.
The latter depends on the isospin projection $T_z$
or, equivalently, on the nucleon numbers $N$ and $Z$,
since it limits the allowed values of isospin through $T\geq|T_z|$.
For two neutrons and two protons, $T_z=0$,
the favoured supermultiplet is $[\bar h]=[4]$ of ${\rm U}_L(6)$,
which corresponds to $[\bar h']=[1,1,1,1]$ of ${\rm U}_{ST}(4)$
or to $(\lambda',\mu',\nu')=(0,0,0)$ of ${\rm SU}_{ST}(4)$.
For three neutrons and one proton or its mirror system, $T_z=\pm1$,
the favoured supermultiplet is $[\bar h]=[3,1]$ or $[\bar h']=[2,1,1]$ or $(\lambda',\mu',\nu')=(1,0,1)$.
And for four identical nucleons, $T_z=\pm2$,
it is $[\bar h]=[\bar h']=[2,2]$ or $(\lambda',\mu',\nu')=(0,2,0)$.
Retaining only the states contained in the favoured supermultiplets,
we find the spectra shown in the right panel of figure~\ref{f_quad1} for $T_z=0$, $\pm1$ and $\pm2$.
Finally, each irreducible representation $(\lambda,\mu)$
corresponds to one or several rotational bands,
as shown in figure~\ref{f_quad1c}
for the case of two neutrons and two protons ($T_z=0$) in the $sd$ shell.

Elliott's SU(3) model contains the correct ingredients
to describe quadrupole-deformed states in the context of the spherical shell model.
The quadrupole interaction implies the orbital reduction~(\ref{e_clasq})
and represents an example of dynamical symmetry breaking.
The degeneracy of states belonging to a Wigner supermultiplet
is lifted dynamically by the quadrupole interaction.
Elliott's SU(3) model gives rise to a rotational classification of states
through mixing of spherical configurations
and shows how deformed nuclear shapes
may emerge out of the spherical shell model.
Elliott's work therefore establishes a link
between the spherical nuclear shell model~\cite{Mayer49,Jensen49}
and the geometric collective model~\cite{Rainwater50,Bohr52,Bohr53},
which originally existed as separate views of the nucleus.

While Elliott's SU(3) model provides a natural explanation of nuclear rotations,
it does so by assuming Wigner's SU(4) symmetry,
which is known to be strongly broken in most but the lightest nuclei.
This raises the following question:
How can rotational phenomena in nuclei be understood
starting from the $jj$-coupling scheme that applies to nearly all nuclei?
Over the years several schemes have been proposed
with the aim of transposing the SU(3) scheme
to those modified situations.
One such modification was suggested by Zuker~{\it et al.}~\cite{Zuker95}
under the name of quasi-SU(3),
invoking the similarities of matrix elements of the quadrupole operator
in the $jj$- and $LS$-coupling schemes.
In a recent study~\cite{Zuker15}
quasi-SU(3) was shown to provide a natural scheme
in which to describe rotational motion
when the dominant single-particle levels satisfy a $\Delta j=2$ condition
and furthermore to give a framework in which to understand
why most deformed nuclei have prolate shapes.

One of the most successful ways
to extend the applications of the SU(3) scheme to heavier nuclei
makes use of the concept of pseudo-spin symmetry,
which can be explained by considering
the single-particle part of the hamiltonian~(\ref{e_hamsm}).
For $\zeta_{\ell\ell}=\zeta_{\ell s}=0$
it displays the degeneracies associated with the U(3) symmetry
of the three-dimensional isotropic harmonic oscillator.
For general values $\zeta_{\ell\ell}\neq0$ and $\zeta_{\ell s}\neq0$
this U(3) symmetry is broken,
except for the combination $4\zeta_{\ell\ell}=\zeta_{\ell s}$
when some degree of degeneracy,
associated with a pseudo-spin symmetry,
is restored in the single-particle spectrum.

The existence of nearly degenerate pseudo-spin doublets
in the nuclear mean-field potential
was pointed out a long time ago by Hecht and Adler~\cite{Hecht69}
and by Arima~{\it et al.}~\cite{Arima69}.
These authors noted that,
while $LS$ coupling becomes unacceptable
in medium-mass and heavy nuclei,
pseudo-$LS$ coupling might be a reasonable starting point
since the pseudo-spin--orbit splitting is small.
With this assumption as a premise,
a pseudo-SU(3) model can be constructed~\cite{Ratna73},
similar to Elliott's SU(3) model in $LS$ coupling.
Many years later Ginocchio showed pseudo-spin
to be a symmetry of the Dirac equation
if the scalar and vector potentials
are equal in size but opposite in sign~\cite{Ginocchio97}.

Many applications of the pseudo-SU(3) scheme to heavy deformed nuclei
have been reported in the literature. 
A summary of representative results was reported in Ref.~\cite{Hirsch03},
where results for bands in the even--even nucleus $^{166}$Er
and in the odd-mass nucleus $^{163}$Dy were given
and excellent agreement between theory and experiment
was achieved in both cases.

\begin{figure}
\centering
\includegraphics[width=8.5cm]{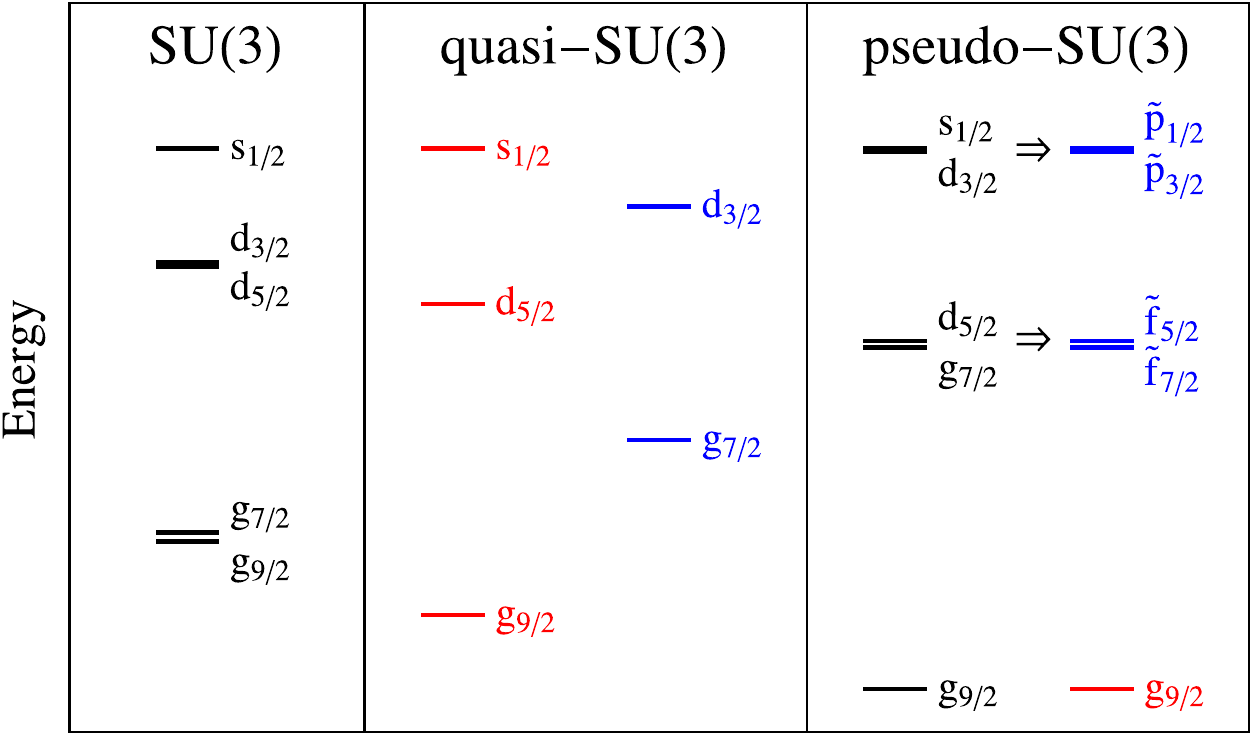}
\caption{The single-particle energies
(for a non-zero orbit--orbit term, $\zeta_{\ell\ell}\neq0$)
in SU(3), quasi-SU(3) and pseudo-SU(3)
for the $N=4$ oscillator shell with the orbital shells $sdg$.
The spin--orbit strength vanishes in SU(3), $\zeta_{\ell s}=0$,
in pseudo-SU(3) $\zeta_{\ell s}=4\zeta_{\ell\ell}$ 
and in quasi-SU(3) a possible choice is $\zeta_{\ell s}\approx2\zeta_{\ell\ell}$.
The single-particle spaces in red and in blue
are assumed to be approximately decoupled.
In pseudo-SU(3) the level degeneracies
can be interpreted in terms of a pseudo-spin symmetry.}
\label{f_qpsu3}
\end{figure}
Figure~\ref{f_qpsu3} provides a schematic illustration
of the various SU(3)-like symmetries
for the $N=4$ ($sdg$) shell of the harmonic oscillator.
On the left-hand side are shown the single-particle energies
in the case of a vanishing spin--orbit term, $\zeta_{\ell s}=0$,
but a non-zero orbit--orbit term, $\zeta_{\ell\ell}\neq0$.
The pseudo-SU(3) symmetry applies if $\zeta_{\ell s}=4\zeta_{\ell\ell}$
while the quasi-SU(3) scenario is valid, for example, for $\zeta_{\ell s}\approx2\zeta_{\ell\ell}$.
At the basis of the quasi-SU(3) and pseudo-SU(3) symmetries
is the assumption of an approximate decoupling of the single-particle spaces
shown in red and blue in the figure.
They can, however, be coupled by non-SU(3) interactions,
{\it e.g.} the pairing interaction.

\subsection{Two oscillator shells}
\label{ss_quad2}
Although not considered originally by Elliott,
it is possible to formulate the SU(3) model for several oscillator shells.
This extension is relatively straightforward
if two consecutive shells of the harmonic oscillator are considered
with the major quantum numbers $N_-\equiv N-1$ and $N_+\equiv N$.
For $N=1,2,3,\dots$ they contain the $s$--$p$, $p$--$sd$, $sd$--$pf$,\dots orbital shells, respectively.
The orbital dimension of this system is $\Omega^2$ with $\Omega\equiv N+1$;
for neutrons and protons the total dimension is $4\Omega^2$
to account for the spin--isospin degrees of freedom. 

A classification of states can be proposed
based on the separation of the orbital and spin--isospin degrees of freedom
which is the analogue of equation~(\ref{e_su4a}),
\begin{equation}
\begin{array}{ccccc}
{\rm U}(4\Omega^2)&\!\!\!\!\supset\!\!\!\!&{\rm U}_L(\Omega^2)&\!\!\!\!\otimes\!\!\!\!&{\rm U}_{ST}(4)\\
\downarrow&&\downarrow&&\downarrow\\[0mm]
[1^n]&&[\bar h]&&[\bar h']
\end{array},
\label{e_su4c}
\end{equation}
where $[\bar h]\equiv[h_1,\dots,h_{\Omega^2}]$
and  $[\bar h']\equiv[h'_1,h'_2,h'_3,h'_4]$.
The generators of the orbital algebra ${\rm U}_L(\Omega^2)$
are the coupled tensors~(\ref{e_tensor}).
In the one-shell case the orbital shells are $\ell=N,N-2,\dots,1$ or 0
while in the two-shell analysis of this subsection
they are $\ell=N,N-1,\dots,0$.

\begin{figure*}
\centering
\includegraphics[height=6.5cm]{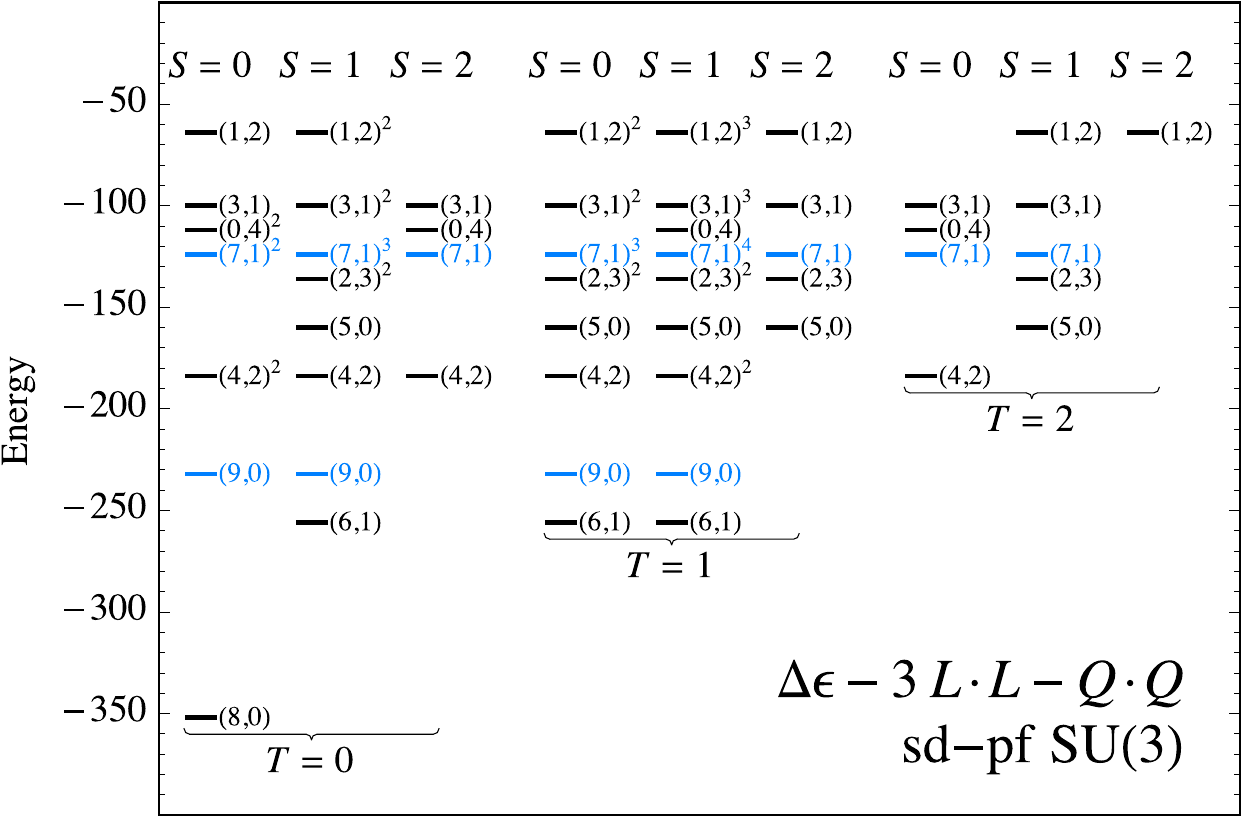}
\includegraphics[height=6.5cm]{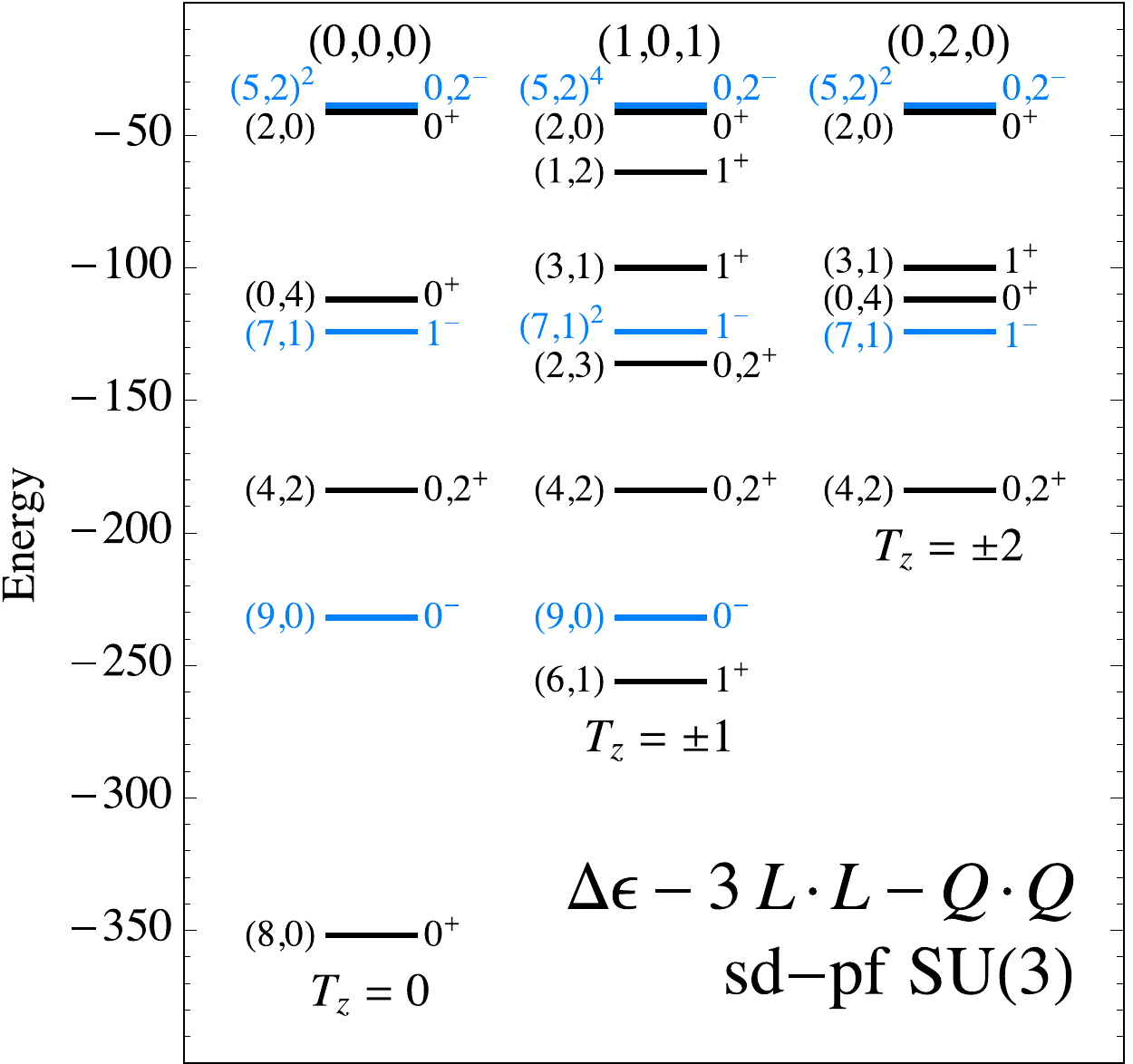}
\caption{Partial eigenspectrum of the hamiltonian
$\Delta\varepsilon(\hat n_+-\hat n_-)+\kappa(3\hat L\cdot\hat L+\hat Q\cdot\hat Q)$
for four nucleons in the $sd$--$pf$ shells with $\Delta\varepsilon=-200\kappa$.
Levels of positive (negative) parity are black (blue).
Left panel:
Levels are labelled by the SU(3) quantum numbers $(\lambda,\mu)$
and values of the total spin $S$ and the total isospin $T$ are also indicated.
If an irreducible representation $(\lambda,\mu)$
occurs $k$ times for a given $S$ and $T$,
this is indicated by a superscript as $(\lambda,\mu)^k$.
Right panel:
Only levels in the favoured supermultiplets are shown,
which are $(0,0,0)$ for $T_z=0$,
$(1,0,1)$ for $T_z=\pm1$
and $(0,2,0)$ for $T_z=\pm2$.
Levels are labelled
by the SU(3) quantum numbers $(\lambda,\mu)$ on the left
and by the projections $K$ and parity $\pi$ on the right.
The supermultiplet labels $(\lambda',\mu',\nu')$
and the isospin projection $T_z$ are also indicated,
and all levels have $S=0$.}
\label{f_quad2}
\end{figure*}
For this two-shell system the following orbital classification can be proposed:
\begin{align}
&\begin{array}{ccccccccc}
{\rm U}(\Omega^2)&\!\!\!\!\supset\!\!\!\!&
{\rm U}(\Gamma_-)&\!\!\!\!\otimes\!\!\!\!&{\rm U}(\Gamma_+)&\!\!\!\!\supset\!\!\!\!&
{\rm U}_-(3)&\!\!\!\!\otimes\!\!\!\!&{\rm U}_+(3)\\
\downarrow&&\downarrow&&\downarrow&&\downarrow&&\downarrow\\[0mm]
[\bar h]&&[\bar h_-]&&[\bar h_+]&&[\bar h''_-]&&[\bar h''_+]
\end{array}
\nonumber\\\nonumber\\[-10pt]
&\begin{array}{cccccccc}
\supset\!\!\!\!&{\rm SU}_-(3)&\!\!\!\!\otimes\!\!\!\!&{\rm SU}_+(3)&\!\!\!\!\supset\!\!\!\!&
{\rm SU}(3)&\!\!\!\!\supset\!\!\!\!&{\rm SO}(3)\\
&\downarrow&&\downarrow&&\downarrow&&\downarrow\\[0mm]
&(\lambda_-,\mu_-)&&(\lambda_+,\mu_+)&&(\lambda,\mu)&K&L
\end{array},
\label{e_clasq2}
\end{align}
where the subscript `$L$' is omitted from the orbital algebras for simplicity's sake
and the notations $\Gamma_-$ and $\Gamma_+$ are introduced
for the orbital dimensions of the lower and upper shells, respectively.
Note that $\Gamma_-+\Gamma_+=\Omega^2$.
The orbital classification~(\ref{e_clasq2})
implies a reduction from the total algebra ${\rm U}(\Omega^2)$
to the product algebra ${\rm U}(\Gamma_-)\otimes{\rm U}(\Gamma_+)$,
where the algebras ${\rm U}(\Gamma_\pm)$ are associated
with the separate oscillator shells with major quantum numbers $N_\pm$.
This requires the knowledge of the branching rule for
${\rm U}(\Omega^2)\supset{\rm U}(\Gamma_-)\otimes{\rm U}(\Gamma_+)$,
which is explained in the appendix.
The determination of the ${\rm SU}_\pm(3)$ labels $(\lambda_\pm,\mu_\pm)$
proceeds as in section~\ref{ss_quad1}
while the labels $(\lambda,\mu)$ associated with SU(3)
follow from the standard multiplication of Young diagrams in U(3)~\cite{Hamermesh62}.

The following hamiltonian can be proposed
for the two-shell SU(3) model:
\begin{equation*}
\varepsilon_-\hat n_-+
\varepsilon_+\hat n_++
\kappa_-\hat Q_-\cdot\hat Q_-+
\kappa_+\hat Q_+\cdot\hat Q_++
\kappa\hat Q\cdot\hat Q+
\kappa'\hat L\cdot\hat L,
\end{equation*}
where $\hat n_\pm$, $\hat L_{\pm,\mu}$ and $\hat Q_{\pm,\mu}$
are the number, orbital angular momentum and quadrupole operators
for the oscillator shells with major quantum numbers $N_\pm$,
and $\hat L_\mu$ and $\hat Q_\mu$ are summed operators,
\begin{equation*}
\hat L_\mu\equiv\hat L_{-,\mu}+\hat L_{+,\mu},
\quad
\hat Q_\mu\equiv\hat Q_{-,\mu}+\hat Q_{+,\mu}.
\end{equation*}
This hamiltonian has the desired flexibility
since it allows an energy difference $\Delta\varepsilon\equiv\varepsilon_+-\varepsilon_-$
between the lower and upper shell,
possibly different strengths
of the quadrupole interactions in the two shells
and a rotational $\hat L\cdot\hat L$ term.
An essentially equivalent hamiltonian
can be written in terms of Casimir operators,
\begin{align}
\hat H={}&
%g\hat C_2[{\rm U}(\Omega^2)]+
\varepsilon_-\hat C_1[{\rm U}(\Gamma_-)]+
\varepsilon_+\hat C_1[{\rm U}(\Gamma_+)]+
\kappa_-\hat C_2[{\rm SU}_-(3)]
\label{e_ham2su3}\\&+
\kappa_+\hat C_2[{\rm SU}_+(3)]+
\kappa\hat C_2[{\rm SU}(3)]+
\kappa'\hat C_2[{\rm SO}(3)].
\nonumber
\end{align}
Since all Casimir operators
are associated with the single chain~(\ref{e_clasq2}) of nested algebras,
this hamiltonian has a dynamical symmetry
and is therefore analytically solvable.
The operator $\hat C_2[{\rm U}(\Omega^2)]$
can be added to the hamiltonian~(\ref{e_ham2su3})
to achieve the separation of supermultiplets,
in accordance with the discussion of section~\ref{s_wigner}.

An example eigenspectrum of the hamiltonian~(\ref{e_ham2su3})
is shown in figure~\ref{f_quad2} for four nucleons in the $sd$-$pf$ shells
({\it i.e.}, $n=4$, $\Gamma_-=6$ and $\Gamma_+=10$).
The spectrum contains the four-nucleon states of the $sd$ shell
(levels in black, identical to those in figure~\ref{f_quad1})
and many additional states that correspond to excitations
of nucleons from the $sd$ to $pf$ shell.
For the choice of parameters in figure~\ref{f_quad2},
$\Delta\varepsilon=-200\kappa$, $\kappa<0$,
the latter excitations occur at higher energy
and only those corresponding to excitations
of one nucleon from the $sd$ to $pf$ shell are shown 
(levels in blue, of negative parity).
The states shown in the left panel of figure~\ref{f_quad2}
belong to the five possible supermultiplets
labelled by $[\bar h]$ or $[\bar h']$ in equation~(\ref{e_su4c}).
With the same assumption as in the single-shell case,
retaining only the states contained in the favoured supermultiplets,
we find the spectra shown in the right panel of figure~\ref{f_quad2} for $T_z=0$, $\pm1$ and $\pm2$.
Some of the states have spurious components,
as discussed in the example that follows.

Let us further illustrate the two-shell SU(3) model with an application to $^{20}$Ne.
In its ground-state configuration this nucleus has four nucleons in the $sd$ shell.
Its observed negative-parity levels presumably result
from excitations of nucleons from the $p$ to $sd$ shell
and a reasonable {\it ansatz} therefore
is to consider $n=16$ nucleons in the $p$--$sd$ shells,
which implies $N_-=1$ and $N_+=2$,
and ${\rm U}(\Omega^2)={\rm U}(9)$,
${\rm U}(\Gamma_-)={\rm U}(3)$
and ${\rm U}(\Gamma_+)={\rm U}(6)$.

The size of the model space for 16 nucleons in the $p$--$sd$ shells
is given by the dimension of the irreducible representation $[1^{16}]$ of U(36),
which, including all magnetic substates in angular momentum and isospin,
is 7~307~872~110.
This dimension can be reduced
by considering a specific magnetic substate ({\it e.g.}, $M_J=M_T=0$)
but the resulting model space will still be huge.
To simplify, we assume in the following
that all low-energy states belong
to the favoured supermultiplet $[\bar h]=[4,4,4,4]$ of ${\rm U}_L(9)$,
which corresponds to $[\bar h']=[4,4,4,4]$ of ${\rm U}_{ST}(4)$
or to $(\lambda',\mu',\nu')=(0,0,0)$ of ${\rm SU}_{ST}(4)$.
This implies that $S=T=0$ 
and therefore that the orbital and total angular momenta are equal, $L=J$.

The next step is the branching rule ${\rm U}(9)\supset{\rm U}(3)\otimes{\rm U}(6)$
for the irreducible representation $[\bar h]=[4,4,4,4]$ of U(9).
The latter has the dimension 1~646~568
and contains 35 product representations $[\bar h_-]\times[\bar h_+]$
of ${\rm U}(3)\otimes{\rm U}(6)$.
Most of them are of no importance for the low-energy spectrum of $^{20}$Ne.
For positive $\Delta\varepsilon$
the lowest-energy product representation is $[\bar h_-]\times[\bar h_+]=[4,4,4]\times[4]$,
which corresponds to twelve nucleons in the $p$ shell
and four in the $sd$ shell.
The next product representation of ${\rm U}(3)\otimes{\rm U}(6)$
contained in $[\bar h]=[4,4,4,4]$ of U(9)
is $[\bar h_-]\times[\bar h_+]=[4,4,3]\times[4,1]$,
and corresponds to the excitation of one nucleon from the $p$ to $sd$ shell.
This correspondence is unique, that is,
a one-particle--one-hole excitation necessarily belongs to $[4,4,3]\times[4,1]$.
It is possible to continue in this fashion for excitations involving more nucleons
but, in view of spurious centre-of-mass components (see below),
this makes little sense
and we limit the discussion here to cross-shell excitations of at most one nucleon.

The ${\rm SU}_-(3)$ labels $(\lambda_-,\mu_-)$ of the lower ($p$) shell are trivially obtained
since they are the differences of the $[\bar h_-]$ labels,
$\lambda_-=h_{-,1}-h_{-,2}$ and $\mu_-=h_{-,2}-h_{-,3}$.
They are therefore $(\lambda_-,\mu_-)=(0,0)$ for twelve nucleons in the $p$ shell
and $(\lambda_-,\mu_-)=(0,1)$ for eleven nucleons (or one hole) in the $p$ shell.
The ${\rm SU}_+(3)$ labels $(\lambda_+,\mu_+)$ for four and five nucleons, respectively,
in the upper ($sd$) shell
follow from the branching rules for ${\rm U}(6)\supset{\rm SU}(3)$,
which can be taken from Elliott~\cite{Elliott58} (see also the appendix),
\begin{align*}
[4]\mapsto{}&(8,0)+(4,2)+(0,4)+(2,0),
\\
[4,1]\mapsto{}&(8,1)+(6,2)+(4,3)+(5,1)+(2,4)+(3,2)
\\&
+(4,0)+(1,3)+(2,1)+(0,2).
\end{align*}

The final step requires the outer multiplication
$(\lambda_-,\mu_-)\times(\lambda_+,\mu_+)$,
which yields the SU(3) labels $(\lambda,\mu)$,
\begin{align*}
(0,0)\times[(8,0)+(4,2)+\cdots]={}&
(8,0)+(4,2)+\cdots,
\\
(0,1)\times[(8,1)+(6,2)+\cdots]={}&
(8,2)+(9,0)
\\&
+(6,3)+(7,1)+\cdots,
\end{align*}
where only the most important SU(3) irreducible representations
of each series are listed.
This completes the discussion of all labels
necessary to determine the low-energy eigenspectrum
of the hamiltonian~(\ref{e_ham2su3}) applied to $^{20}$Ne.

\begin{figure*}
\centering
\includegraphics[width=8.5cm]{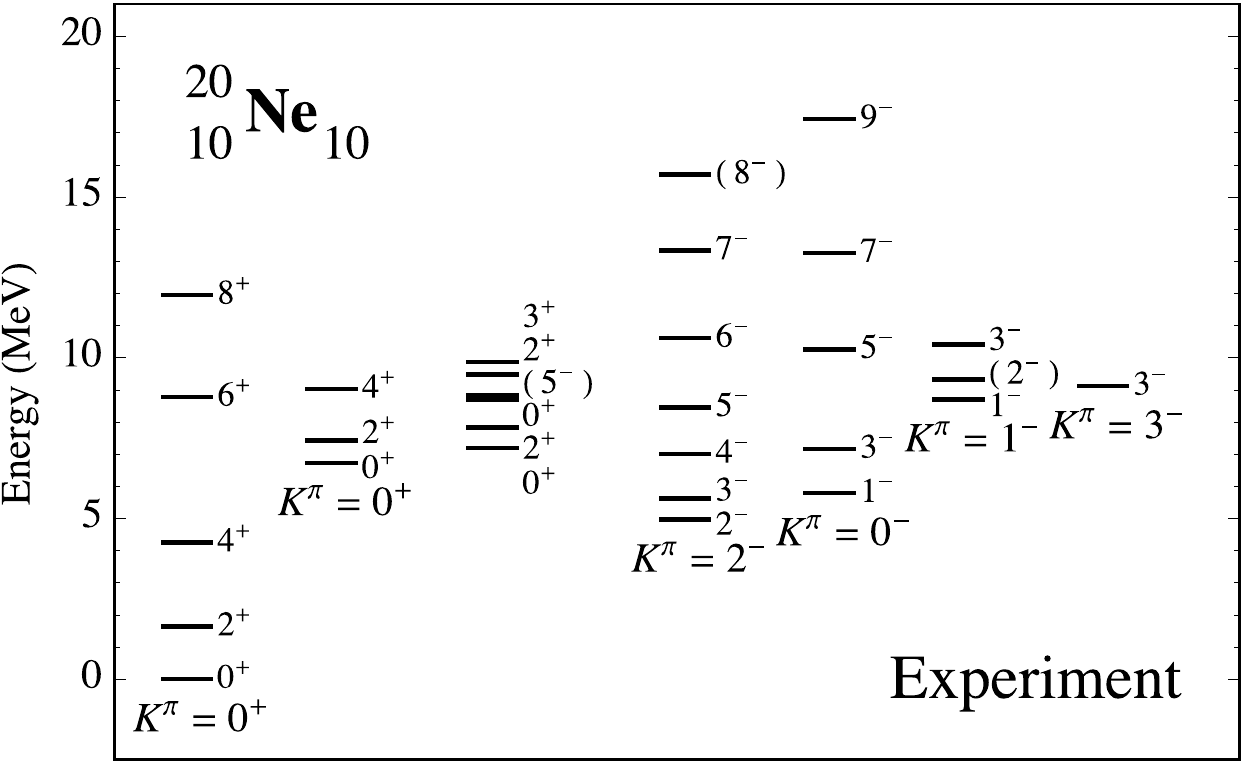}
\includegraphics[width=8.5cm]{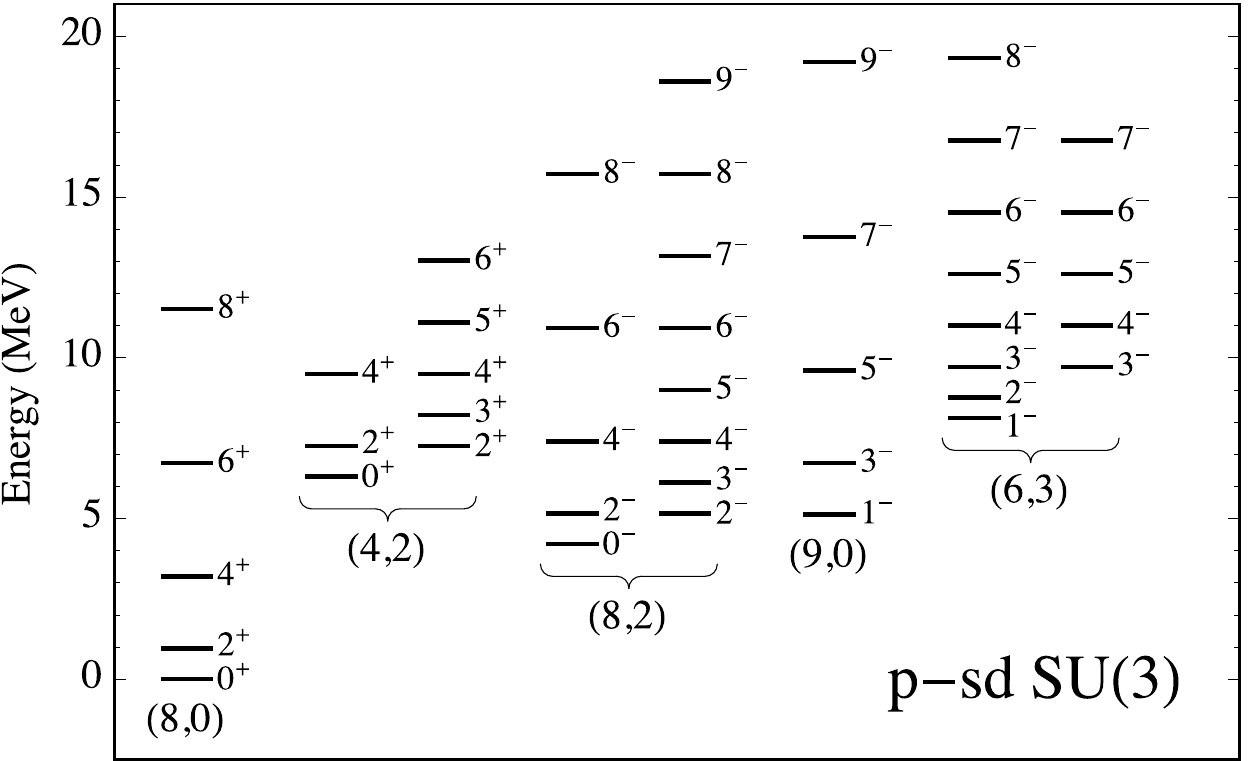}
\caption{Observed energy spectrum of $^{20}$Ne
(taken from NNDC~\cite{NNDC})
compared with the eigenspectrum of a two-shell SU(3) hamiltonian.
Left panel: All observed levels up to an energy of 10~MeV are shown.
Levels that can be associated with a theoretical counterpart
are arranged as on the right panel
and additional levels are shown in the third column.
Uncertain spin--parity assignments are given in brackets
and bands are tentatively assigned a $K$ quantum number, in accordance with their E2 decay properties.
Right panel: The two-shell SU(3) hamiltonian~(\ref{e_ham2su3}) is used
for 16 nucleons in the $p$--$sd$ shells
with parameters (in MeV)
$\Delta\varepsilon\equiv\varepsilon_+-\varepsilon_-=7.60$,
$\kappa_-=\kappa_+=-0.025$, $\kappa=-0.05$ and $\kappa'=-0.16$.
Entire rotational bands are shown and labelled by the SU(3) quantum numbers $(\lambda,\mu)$.}
\label{f_ne20}
\end{figure*}
In figure~\ref{f_ne20} the eigenspectrum of the hamiltonian~(\ref{e_ham2su3}),
with parameters (in MeV)
$\Delta\varepsilon=7.60$,
$\kappa_-=\kappa_+=-0.05$,
$\kappa=-0.10$
and $\kappa'=-0.16$,
is compared with the observed spectrum of $^{20}$Ne,
taken from NNDC~\cite{NNDC}.
From the levels shown in the figure it is not possible to determine separately
the parameters $\Delta\varepsilon$ and $\kappa_-$,
and therefore $\kappa_-=\kappa_+$ is taken,
which corresponds to equal quadrupole strengths in the $p$ and $sd$ shells.
The assumption of different strengths, $\kappa_-\neq\kappa_+$,
has no impact on the quality of the fit
and leads to a (slightly) different value of $\Delta\varepsilon$.
All observed levels up to an energy of 10~MeV are shown in figure~\ref{f_ne20}
and most can be associated with a theoretical counterpart.
Among the additional observed levels (shown in the third column of the left panel)
several could be members of the collective $\gamma$-vibrational $K^\pi=2^+$ band
or be interpreted as cross-shell excitations of several nucleons.
On the theoretical side, a $K^\pi=2^+$ band is calculated
in the $(\lambda,\mu)=(4,2)$ irreducible representation of SU(3),
which, surprisingly, has not been established experimentally.
More conspicuous is the presence in the theory of an additional $K^\pi=0^-$ band
in the $(\lambda,\mu)=(8,2)$ irreducible representation of SU(3),
which is not seen experimentally.

If more than one oscillator shell is considered in the model space,
care must be taken to eliminate spurious centre-of-mass components
from the calculated states.
If the model space consists of entire shells of the harmonic oscillator,
an exact procedure exists for doing so,
based on the action of the centre-of-mass operator $\bar R$
on the ground state of a given nucleus,
and the elimination from the excitation spectrum
of the states so created~\cite{Elliott55}.
The method is particularly attractive in the SU(3) scheme
since the operator $\bar R$ has a $[1,0,0]$ tensor character under U(3).
Therefore, if the ground state belongs
to a certain irreducible representation $(\lambda,\mu)$ of SU(3),
spurious states follow from the repeated multiplication of $(\lambda,\mu)$ with $(1,0)$.
In the example of $^{20}$Ne
the single action of $\bar R$ on the ground state
leads to spurious excitations with SU(3) character (9,0) and (7,1) since
\begin{equation*}
(8,0)\times(1,0)=(9,0)+(7,1).
\end{equation*}
As one of the irreducible representations, (9,0),
occurs in the low-energy spectrum,
we must prove that it is not spurious.

It should be noted that the technique
for eliminating spurious centre-of-mass components
cannot be accommodated in the classification~(\ref{e_clasq2})
because the action of the operator $\bar R$ creates additional excitations
outside the model space.
In the example of $^{20}$Ne,
which has a ground-state configuration with a completely filled $p$ shell 
and four nucleons in the $sd$ shell,
$\bar R$ excites nucleons not only from the $p$ to $sd$ shell
[excitations that are included in the classification~(\ref{e_clasq2})]
but also from the $sd$ to $pf$ shell (excitations that are not).
For a correct elimination of spurious centre-of-mass components
from all one-particle--one-hole excitations,
it is therefore necessary to consider the $p$, $sd$ and $pf$ shells---a
straightforward extension of the two-shell system considered so far.
All previous results remain valid
and, in addition, the spectrum should be complemented
with excitations of one nucleon from the $sd$ to $pf$ shell.
The lowest-energy excitations of the latter type have SU(3) character
\begin{equation*}
(6,0)\times(3,0)=(9,0)+(7,1)+\cdots,
\end{equation*}
corresponding to three nucleons in the $sd$ and one nucleon in the $pf$ shell.
We conclude that in the three-shell $p$--$sd$--$pf$ system
{\em two} irreducible representations (9,0) occur
among the low-energy one-particle--one-hole excitations.
One is spurious and must be eliminated;
the other is physical and is included in figure~\ref{f_ne20}.

These results are obtained under the assumption
that all observed states in $^{20}$Ne 
belong to the favoured supermultiplet $[\bar h]=[4,4,4,4]$ of ${\rm U}_L(9)$.
While the hypothesis of maximal spatial symmetry
might be acceptable for the single-shell SU(3) model,
it is more questionable in the case of several shells.
In Ref.~\cite{Cseh15} it is argued
that this hypothesis is related to quartet clustering
and that the observed levels in $^{20}$Ne
should therefore be interpreted as cluster states.
The combined classifications~(\ref{e_su4c}) and~(\ref{e_clasq2}),
suitably extended to the appropriate multi-shell scenario,
enable one to study the interplay between
the maximization of spatial symmetry on the one hand
and cross-shell excitations on the other,
in order to verify the validity of the cluster interpretation.

\subsection{Many oscillator shells}
\label{ss_quadm}
It is instructive to abandon for a moment the formalism of second quantization
and return to the representation of operators in coordinate and momentum space.
As mentioned already, U(3) generators can be represented in a spherical basis
and the components of the quadrupole operator in this basis are~\cite{Elliott58}
\begin{equation*}
\hat Q_\mu=
\sqrt{\frac32}
\left(\sum_{k=1}^A(\bar r_k\times\bar r_k)^{(2)}_\mu/b^2+
b^2\sum_{k=1}^A(\bar p_k\times\bar p_k)^{(2)}_\mu/\hbar^2\right),
\end{equation*}
where $b$ is the length parameter of the harmonic oscillator,
$b=\sqrt{\hbar/m_{\rm n}\omega}$.
Because of its dependence on the position vectors $\bar r_k$
and the momentum vectors $\bar p_k$ of the nucleons,
matrix elements of $\hat Q_\mu$ {\em between} oscillator shells vanish,
and it is precisely this property
which gives rise to the algebraic structure of the SU(3) model.
The components of the physical quadrupole operator, on the other hand, are
\begin{equation*}
\hat Q'_\mu=
\sqrt{6}
\sum_{k=1}^A (\bar r_k\times\bar r_k)^{(2)}_\mu/b^2.
\end{equation*}
In a single oscillator shell,
it makes no difference whether the algebraic quadrupole operator $\hat Q_\mu$
or the physical quadrupole operator $\hat Q'_\mu$ is used
(matrix elements of both operators are identical due to the virial theorem).
Because of parity, this statement remains valid for two shells $N-1$ and $N$.
Differences arise however, if more than two shells are considered
because $\hat Q'_\mu$ has non-zero $\Delta N=\pm2$ matrix elements.
These are important
because they are responsible for the full quadrupole collectivity in nuclei.
For the sake of constructing a closed algebra,
cross-shell correlations are thus lost from the SU(3) model.

By embedding SU(3) into the symplectic algebra ${\rm Sp}(6,R)$,
it is possible to accommodate cross-shell effects~\cite{Rosensteel77,Rosensteel80}.
(We follow Gilmore's notation of the Lie algebras of the classical groups~\cite{Gilmore74}.)
The following relation between the algebraic and physical quadrupole operators
illustrates the essence of this idea~\cite{Draayer93}:
\begin{equation*}
\hat Q'_\mu=\hat Q_\mu
+\sqrt{\frac32}
(\hat B^2_{+,\mu}+\hat B^2_{-,\mu}),
\end{equation*}
where $\hat B^2_{\pm,\mu}$
are $2\hbar\omega$ raising and lowering operators
with tensor character $\ell=2$.
The ${\rm Sp}(6,R)$ algebra consists of
$\{\hat n,\hat L_\mu,\hat Q_\mu,
\hat B^0_{\pm,0},\hat B^2_{\pm,\mu}\}$,
where $\hat n$ is the number operator,
$\hat L_\mu$ and $\hat Q_\mu$ are the components
of the angular momentum and quadrupole operators forming SU(3),
and $\hat B^\ell_{\pm,\mu}$ are the monopole ($\ell=0$) and quadrupole ($\ell=2$)
$2\hbar\omega$ raising and lowering operators.
These operators close under commutation
and a linear combination of them
corresponds to the physical quadrupole operator $\hat Q'_\mu$.
The ${\rm Sp}(6,R)$ model contains all the necessary ingredients
for the description of quadrupole-deformed states
in the context of the spherical shell model
with enhanced collectivity due to cross-shell excitations.

\section{A shell-model classification for octupole deformation}
\label{s_octu}
The question arises whether a generic classification exists
for octupole-deformed shell-model states,
similar to the one constructed by Elliott for quadrupole deformation.
The minimal realization in this case should be
in terms of the three components $\hat L_\mu$ of the angular momentum operator
and the seven components $\hat O_\mu$ of the octupole operator.
As the octupole operators should be of negative parity,
at least two harmonic-oscillator shells
are required for this minimal realization.

We show that a symmetry of this type indeed exists,
by following the treatment of section~\ref{ss_quad2}
and considering two consecutive oscillator shells
with the major quantum numbers $N_-\equiv N-1$ and $N_+\equiv N$.

The octupole classification discussed in this section
is inspired by studies in the context of the interacting boson model~\cite{Iachello87},
in particular the U(16) $spdf$ version of it,
which deals with the quantization of asymmetric shapes in nuclei~\cite{Engel85}.
This model, which has been extensively studied by Kusnezov~\cite{Kusnezov89,Kusnezov90},
has also been extended to odd-mass nuclei~\cite{Engel87}.
Our approach is different, however,
since it is entirely fermionic and takes fully into account the Pauli principle.
Just as Elliott's SU(3) symmetry is not to be confused
with the SU(3) limit of the interacting boson model,
the classification proposed here differs from the limits of the U(16) model.
Also, we propose a symmetry treatment,
which, like Elliott's, is generic and applies to any two consecutive major oscillator shells.

\subsection{A solvable dipole--octupole hamiltonian}
\label{s_octu2}
The starting point of the analysis
is based on the separation of the orbital and spin--isospin degrees of freedom,
identical to equation~(\ref{e_su4c}).
The generators of the orbital algebra ${\rm U}_L(\Omega^2)$
can be written as coupled tensors~(\ref{e_tensor})
but also as {\em double} tensors
\begin{align}
\hat G^{(\lambda_1\lambda_2)}_{\mu_1\mu_2}\equiv{}&
\sum_{\ell_1\ell_2}\sum_{\lambda\mu}
(\lambda_1\mu_1\,\lambda_2\mu_2|\lambda\mu)
\nonumber\\&\times
\left[\!\!\begin{array}{ccc}
{\frac12}N&{\frac12}N&\ell_1\\[0.5ex]
{\frac12}N&{\frac12}N&\ell_2\\[0.5ex]
\lambda_1&\lambda_2&\lambda
\end{array}\!\!\right]
\hat G^{(\lambda)}_\mu(\ell_1\ell_2).
\label{e_dtensor}
\end{align}
The orbital shells in the two-shell analysis of this section are $\ell=N,N-1,\dots,0$,
as shall be implicitly assumed henceforth in all summations.

It is not {\it a priori} clear what the physical meaning of the double tensors is.
Nevertheless, as will be shown below,
a linear combination of the operators~(\ref{e_dtensor})
with $(\lambda_1,\lambda_2)=(3,0)$ and $(0,3)$
approximately corresponds to the octupole operator $\hat O_\mu$.
Also, it is the double-tensor character of the generators
that enables the definition of a product of unitary algebras
in the classification proposed below,
leading to the occurrence of parity doublets in the spectrum.

For nucleons occupying two major shells $N-1$ and $N$,
the following orbital classification can be proposed:
\begin{align}
&\begin{array}{cccccccccc}
{\rm U}(\Omega^2)&\!\!\!\!\supset\!\!\!\!&
{\rm U}_a(\Omega)&\!\!\!\!\otimes\!\!\!\!&{\rm U}_b(\Omega)&\!\!\!\!\supset\!\!\!\!&
{\rm SOp}_a(\Omega)&\!\!\!\!\otimes\!\!\!\!&{\rm SOp}_b(\Omega)\\
\downarrow&&\downarrow&&\downarrow&&\downarrow&&\downarrow\\[0mm]
[\bar h]&&[\bar h_a]&&[\bar h_b]&&\langle\bar\omega_a\rangle&&\langle\bar\omega_b\rangle
\end{array}
\nonumber\\\nonumber\\[-10pt]
&\begin{array}{cccccccc}
\supset\!\!\!\!&{\rm SOp}_+(\Omega)&\!\!\!\!\supset\!\!\!\!&{\rm SO}_+(3)\\
&\downarrow&&\downarrow\\[0mm]
&\langle\bar\omega_+\rangle&&L
\end{array},
\label{e_claso1}
\end{align}
with $\Omega\equiv N+1$
and where the subscript `$L$' again is omitted from the orbital algebras for simplicity's sake.
The (non-standard) abbreviation ${\rm SOp}(\Omega)$ refers
to the (unitary) symplectic algebra ${\rm Sp}(\Omega)$ if $\Omega$ is even
and to the orthogonal algebra ${\rm SO}(\Omega)$ if $\Omega$ is odd.
The labels underneath the algebras are explained below.

The classification~(\ref{e_claso1}) follows
from the commutator property of the double tensors
$\hat G^{(\lambda_1\lambda_2)}_{\mu_1\mu_2}$, which reads
\begin{align}
&[\hat G^{(\lambda_1\lambda_2)}_{\mu_1\mu_2},\hat G^{(\lambda_3\lambda_4)}_{\mu_3\mu_4}]=
{\frac12}\hat\lambda_1\hat\lambda_2\hat\lambda_3\hat\lambda_4
\sum_{\lambda_{13}\mu_{13}}\sum_{\lambda_{24}\mu_{24}}
\label{e_comuom}\\&\times
\left[(-)^{\lambda_{13}+\lambda_{24}}-(-)^{\lambda_1+\lambda_2+\lambda_3+\lambda_4}\right]
\nonumber\\&\times
(\lambda_1\mu_1\,\lambda_3\mu_3|\lambda_{13}\mu_{13})
(\lambda_2\mu_2\,\lambda_4\mu_4|\lambda_{24}\mu_{24})
\nonumber\\&\times
\left\{\!\!\begin{array}{ccc}
\lambda_1&\lambda_3&\lambda_{13}\\
{\frac12}N&{\frac12}N&{\frac12}N
\end{array}\!\!\right\}
\left\{\!\!\begin{array}{ccc}
\lambda_2&\lambda_4&\lambda_{24}\\
{\frac12}N&{\frac12}N&{\frac12}N
\end{array}\!\!\right\}
\hat G^{(\lambda_{13}\lambda_{24})}_{\mu_{13}\mu_{24}}.
\nonumber
\end{align}
The derivation of this commutator relation
requires the expansion of the double tensors~(\ref{e_dtensor})
in terms of uncoupled generators.
It makes use of the anti-commutators~(\ref{e_anticom})
and of summation properties of Clebsch--Gordan coefficients, six-$j$  and nine-$j$ symbols~\cite{Talmi93}.
The corresponding relation for bosons (without the coefficient 1/2)
is given by Kusnezov for $N=3$
in connection with the $spdf$ interacting boson model~\cite{Kusnezov88}.
Many properties concerning the various octupole classifications
can be derived from equation~(\ref{e_comuom}).

From the commutator property~(\ref{e_comuom}) it immediately follows that
$\hat G^{(0\lambda_1)}_{0\mu_1}$ and $\hat G^{(\lambda_20)}_{\mu_20}$ commute,
\begin{equation*}
[\hat G^{(0\lambda_1)}_{0\mu_1},\hat G^{(\lambda_20)}_{\mu_20}]=0.
\end{equation*}
It also follows that the operators $\hat G^{(0\lambda)}_{0\mu}$ and $\hat G^{(\lambda0)}_{\mu0}$
separately close under commutation since
\begin{align}
&[\hat G^{(0\lambda_1)}_{0\mu_1},\hat G^{(0\lambda_2)}_{0\mu_2}]=
{\frac12}\hat\lambda_1\hat\lambda_2
\sum_{\lambda\mu}
\left[(-)^\lambda-(-)^{\lambda_1+\lambda_2}\right]
\nonumber\\&\quad\times
(\lambda_1\mu_1\,\lambda_2\mu_2|\lambda\mu)
\left\{\!\!\begin{array}{ccc}
\lambda_1&\lambda_2&\lambda\\
{\frac12}N&{\frac12}N&{\frac12}N
\end{array}\!\!\right\}
\hat G^{(0\lambda)}_{0\mu},
\label{e_comuoma}
\end{align}
and
\begin{align}
&[\hat G^{(\lambda_10)}_{\mu_10},\hat G^{(\lambda_20)}_{\mu_20}]=
{\frac12}\hat\lambda_1\hat\lambda_2
\sum_{\lambda\mu}
\left[(-)^\lambda-(-)^{\lambda_1+\lambda_2}\right]
\nonumber\\&\quad\times
(\lambda_1\mu_1\,\lambda_2\mu_2|\lambda\mu)
\left\{\!\!\begin{array}{ccc}
\lambda_1&\lambda_2&\lambda\\
{\frac12}N&{\frac12}N&{\frac12}N
\end{array}\!\!\right\}
\hat G^{(\lambda0)}_{\mu0}.
\label{e_comuomb}
\end{align}
This shows that the decomposition from ${\rm U}(\Omega^2)$
into the product algebra ${\rm U}_a(\Omega)\otimes{\rm U}_b(\Omega)$
is achieved by requiring a scalar character
in one of the indices of the double tensor~(\ref{e_dtensor}),
\begin{align*}
{\rm U}_a(\Omega)&=\{\hat G^{(0\lambda)}_{0\mu},\lambda=0,1,\dots,N\},
\\
{\rm U}_b(\Omega)&=\{\hat G^{(\lambda0)}_{\mu0},\lambda=0,1,\dots,N\}.
\end{align*}
The explicit expression of the generators of ${\rm U}_a(\Omega)$ and ${\rm U}_b(\Omega)$
in terms of the tensors~(\ref{e_tensor}) is
\begin{align}
\hat G^{(0\lambda)}_{0\mu}&=
\sum_{\ell\ell'}
\frac{(-)^{N+\lambda+\ell}}{\sqrt{N+1}}
\hat\ell\hat\ell'
\left\{\!\!\!\begin{array}{ccc}
\ell&\lambda&\ell'\\
{\frac12}N&{\frac12}N&{\frac12}N
\end{array}\!\!\!\right\}
\hat G^{(\lambda)}_\mu(\ell\ell'),
\nonumber\\
\hat G^{(\lambda0)}_{\mu0}&=
\sum_{\ell\ell'}
\frac{(-)^{N+\lambda+\ell'}}{\sqrt{N+1}}
\hat\ell\hat\ell'
\left\{\!\!\!\begin{array}{ccc}
\ell&\lambda&\ell'\\
{\frac12}N&{\frac12}N&{\frac12}N
\end{array}\!\!\!\right\}
\hat G^{(\lambda)}_\mu(\ell\ell').
\nonumber\\
\label{e_dtensor2}
\end{align}
The subsequent reduction to ${\rm Sp}(\Omega)$ or ${\rm SO}(\Omega)$ in equation~(\ref{e_claso1})
follows from the restriction to odd-integer $\lambda$,
\begin{align*}
{\rm SOp}_a(\Omega)&=\{\hat G^{(0\lambda)}_{0\mu},\lambda={\rm odd}\},
\\
{\rm SOp}_b(\Omega)&=\{\hat G^{(\lambda0)}_{\mu0},\lambda={\rm odd}\},
\end{align*}
which close under commutation
because of the presence of the phase factor
in equations~(\ref{e_comuoma}) and~(\ref{e_comuomb}).
The algebra ${\rm SOp}_+(\Omega)$
is obtained by adding the generators
of ${\rm SOp}_a(\Omega)$ and ${\rm SOp}_b(\Omega)$,
\begin{equation*}
{\rm SOp}_+(\Omega)=
\{\hat G^{(0\lambda)}_{0\mu}+\hat G^{(\lambda0)}_{\mu0},\lambda={\rm odd}\}.
\end{equation*}
The components of the operator 
\begin{align}
\hat L_\mu&\equiv\left[\frac{N(N+1)^2(N+2)}{3}\right]^{1/2}
\left(\hat G^{(01)}_{0\mu}+\hat G^{(10)}_{\mu0}\right)
\nonumber\\&=
\sum_\ell
\sqrt{\frac{4\ell(\ell+1)(2\ell+1)}{3}}
\hat G^{(1)}_\mu(\ell\ell),
\label{e_ang1}
\end{align}
are the orbital angular momentum generators
and therefore the algebra ${\rm SO}_+(3)$
coincides with the orbital angular momentum algebra.

The algebra ${\rm SOp}_+(\Omega)$ contains, besides $\hat L_\mu$, 
the components of the octupole operator
$\hat G^{(03)}_{0\mu}+\hat G^{(30)}_{\mu0}$;
it is of positive parity, however,
and therefore does not correspond to the $r^3Y_{3\mu}$ operator
of relevance for octupole deformation in nuclei.
To obtain negative-parity operators, we consider the combinations
\begin{equation*}
{\rm SOp}_-(\Omega)=
\{\hat G^{(0\lambda)}_{0\mu}-\hat G^{(\lambda0)}_{\mu0},\lambda={\rm odd}\}.
\end{equation*}
We use the notation
\begin{equation}
\hat T^{(\lambda)}_\mu\equiv\sqrt{8(N+1)}
\left(\hat G^{(0\lambda)}_{0\mu}-\hat G^{(\lambda0)}_{\mu0}\right),
\label{e_sopm2}
\end{equation}
where the coefficient is introduced for later convenience,
and we adopt furthermore the notations $\hat D_\mu\equiv\hat T^{(1)}_\mu$
and $\hat O_\mu\equiv\hat T^{(3)}_\mu$
for the dipole and octupole operators, respectively.

The generators of ${\rm SOp}_a(\Omega)$
commute with those of ${\rm SOp}_b(\Omega)$
but this is not the case for those of ${\rm SOp}_+(\Omega)$ and ${\rm SOp}_-(\Omega)$.
Also, the generators of ${\rm SOp}_\pm(\Omega)$ are parity-conserving
while this is not so for ${\rm SOp}_{a,b}(\Omega)$.

In the orbital classification~(\ref{e_claso1})
appear the ${\rm U}(\Omega^2)$ labels $[\bar h]$
that are carried over from Wigner's supermultiplet model.
The algebras ${\rm U}_a(\Omega)$ and ${\rm U}_b(\Omega)$
are characterized by the Young diagrams $[\bar h_a]$ and $[\bar h_b]$, respectively,
which can have at most $\Omega$ rows.
They follow from the branching rule for
${\rm U}(\Omega^2)\supset{\rm U}(\Omega)\otimes{\rm U}(\Omega)$,
the algorithm of which is explained in the appendix.
The next step involves twice ({\it i.e.}, for $a$ and for $b$)
the branching rule for ${\rm U}(\Omega)\supset{\rm SOp}(\Omega)$,
that is, ${\rm U}(\Omega)\supset{\rm Sp}(\Omega)$ if $\Omega$ is even,
and ${\rm U}(\Omega)\supset{\rm SO}(\Omega)$ if $\Omega$ is odd.
The number of labels for ${\rm Sp}(\Omega)$ is $\Omega/2$
while it is $(\Omega-1)/2$ for ${\rm SO}(\Omega)$~\cite{Iachello06}.
The labels are denoted in equation~(\ref{e_claso1}) as
$\langle\bar\omega\rangle\equiv\langle\omega_1,\omega_2,\dots\rangle$,
notation which is thus used for irreducible representations
of either symplectic or orthogonal algebras.
In the following, the notation $\{\bar\nu\}\equiv\{\nu_1,\nu_2,\dots\}$
shall be reserved for symplectic algebras ($N$ odd)
and $(\bar\upsilon)\equiv(\upsilon_1,\upsilon_2,\dots)$ for orthogonal algebras ($N$ even).
The branching rules for ${\rm U}(\Omega)\supset{\rm Sp}(\Omega)$
and ${\rm U}(\Omega)\supset{\rm SO}(\Omega)$
can be obtained for any $\Omega$ from $S$-function theory for restricted algebras
and they determine the allowed labels
$\langle\bar\omega_a\rangle$ and $\langle\bar\omega_b\rangle$
in the classification~(\ref{e_claso1}).
The next step requires the multiplication
$\langle\bar\omega_a\rangle\times\langle\bar\omega_b\rangle$,
that is, the outer multiplication of irreducible representations
in either ${\rm Sp}(\Omega)$ or ${\rm SO}(\Omega)$.
This is also known for any $\Omega$ from $S$-function theory for restricted algebras
and yields the irreducible representations $\langle\bar\omega_+\rangle$ of ${\rm SOp}_+(\Omega)$.
Finally, the branching rules for
${\rm Sp}(\Omega)\supset{\rm SO}(3)$ or ${\rm SO}(\Omega)\supset{\rm SO}(3)$
determine the allowed orbital angular momenta $L$
and can be obtained from the plethysm of $S$ functions.

Group-theoretical methods related to $S$-function theory
are described, for example, in the book by Wybourne~\cite{Wybourne70}
which also includes tables (by Butler) with outer multiplications of $S$ functions,
expansions of characters of restricted ({\it i.e.}, symplectic or orthogonal) algebras into $S$ functions
and various branching rules.
Many (but not all) cases of interest for the classification~(\ref{e_claso1})
can be obtained from the tables.
The techniques needed for the present application
are described in the appendix.

For low values of $N$,
isomorphisms between orthogonal and symplectic algebras exist,
${\rm Sp}(2)\sim{\rm SO}(3)$ and ${\rm Sp}(4)\sim{\rm SO}(5)$.
Therefore, up to $N=4$ we may choose
to use the orthogonal algebras SO(3) ($N=1$ or 2) and SO(5) ($N=3$ or 4).
For odd values of $N$, spinor representations of the orthogonal algebras
should be employed, the correspondence with the symplectic labels being
\begin{align*}
{\rm Sp}(2)\sim{\rm SO}(3):\quad&\upsilon_1=\frac{\nu_1}{2},
\\
{\rm Sp}(4)\sim{\rm SO}(5):\quad&\upsilon_1=\frac{\nu_1+\nu_2}{2},
\quad
\upsilon_2=\frac{\nu_1-\nu_2}{2}.
\end{align*}

The quadratic Casimir operator of any of the ${\rm SOp}(\Omega)$ algebras
is defined as
\begin{equation}
\hat C_2[{\rm SOp}(\Omega)]=
8(N+1)\sum_{\lambda\,{\rm odd}}\hat{\cal G}^{(\lambda)}\cdot\hat{\cal G}^{(\lambda)},
\label{e_c2sop}
\end{equation}
where $\hat{\cal G}^{(\lambda)}_\mu$ can be
$\hat G^{(0\lambda)}_{0\mu}$ or $\hat G^{(\lambda0)}_{\mu0}$,
or the combinations
$\hat G^{(0\lambda)}_{0\mu}\pm\hat G^{(\lambda0)}_{\mu0}$.
The eigenvalue expressions are known from classical group theory
(see, for example, table~5.1 of Ref.~\cite{Iachello06}).
The eigenvalues of the operator $\hat C_2[{\rm Sp}(\Omega)]$, for even $\Omega$, are
\begin{equation*}
\sum_{i=1}^{\Omega/2}\nu_i(\nu_i+\Omega+2-2i),
\end{equation*}
while those of the operator $\hat C_2[{\rm SO}(\Omega)]$, for odd $\Omega$, are
\begin{equation*}
\sum_{i=1}^{(\Omega-1)/2}\upsilon_i(\upsilon_i+\Omega-2i).
\end{equation*}
The two expressions can be combined by introducing
\begin{equation}
E_N(\bar\omega)\equiv
\sum_{i=1}^{\lfloor\Omega/2\rfloor}
\omega_i(\omega_i+\Omega_N+2-2i),
\label{e_eigsop}
\end{equation}
with $\Omega_N\equiv\Omega=N+1$ for odd $N$ (symplectic algebras)
and  $\Omega_N\equiv\Omega-2=N-1$ for even $N$ (orthogonal algebras).
Casimir operators are only determined up to a proportionality factor
and the coefficient $8(N+1)$ in equation~(\ref{e_c2sop}) is chosen such
that the expectation value of $\hat C_2[{\rm SOp}(\Omega)]$
yields the eigenvalue~(\ref{e_eigsop}).

A definition similar to equation~(\ref{e_c2sop})
holds for the quadratic Casimir operator of the SO(3) algebras,
\begin{equation}
\hat C_2[{\rm SO}_\pm(3)]=
\frac{N(N+1)^2(N+2)}{3}
\hat{\cal G}^{(1)}\cdot\hat{\cal G}^{(1)},
\label{e_c2so3}
\end{equation}
where $\hat{\cal G}^{(1)}_\mu=\hat G^{(01)}_{0\mu}\pm\hat G^{(10)}_{\mu0}$.
The coefficient in equation~(\ref{e_c2so3}) is chosen such
that the expectation value of $\hat C_2[{\rm SO}_+(3)]$
yields the eigenvalue $L(L+1)$.

From the previous results it follows that
\begin{align}
-\sum_{\lambda\,{\rm odd}}\hat T^{(\lambda)}\cdot\hat T^{(\lambda)}={}&
-2\hat C_2[{\rm SOp}_a(\Omega)]
-2\hat C_2[{\rm SOp}_b(\Omega)]
\nonumber\\&
+\hat C_2[{\rm SOp}_+(\Omega)].
\label{e_hamo1}
\end{align}
Since this hamiltonian can be written
as a combination of Casimir operators
belonging to the chain~(\ref{e_claso1}) of nested algebras,
it is solvable with eigenstates
\begin{equation}
|[1^n];([\bar h];[\bar h_a]\langle\bar\omega_a\rangle\times[\bar h_b]\langle\bar\omega_b\rangle;
\langle\bar\omega_+\rangle L)\times[\bar h']ST\rangle,
\label{e_baso1}
\end{equation}
and energy eigenvalues
\begin{equation}
-2E_N(\bar\omega_a)-2E_N(\bar\omega_b)+E_N(\bar\omega_+).
\label{e_eigo1}
\end{equation}
This establishes the result that a hamiltonian
which is a sum over odd, negative-parity tensors is solvable.

\subsection{Parity doublets}
\label{ss_parity}
\begin{figure*}
\centering
\includegraphics[height=6.5cm]{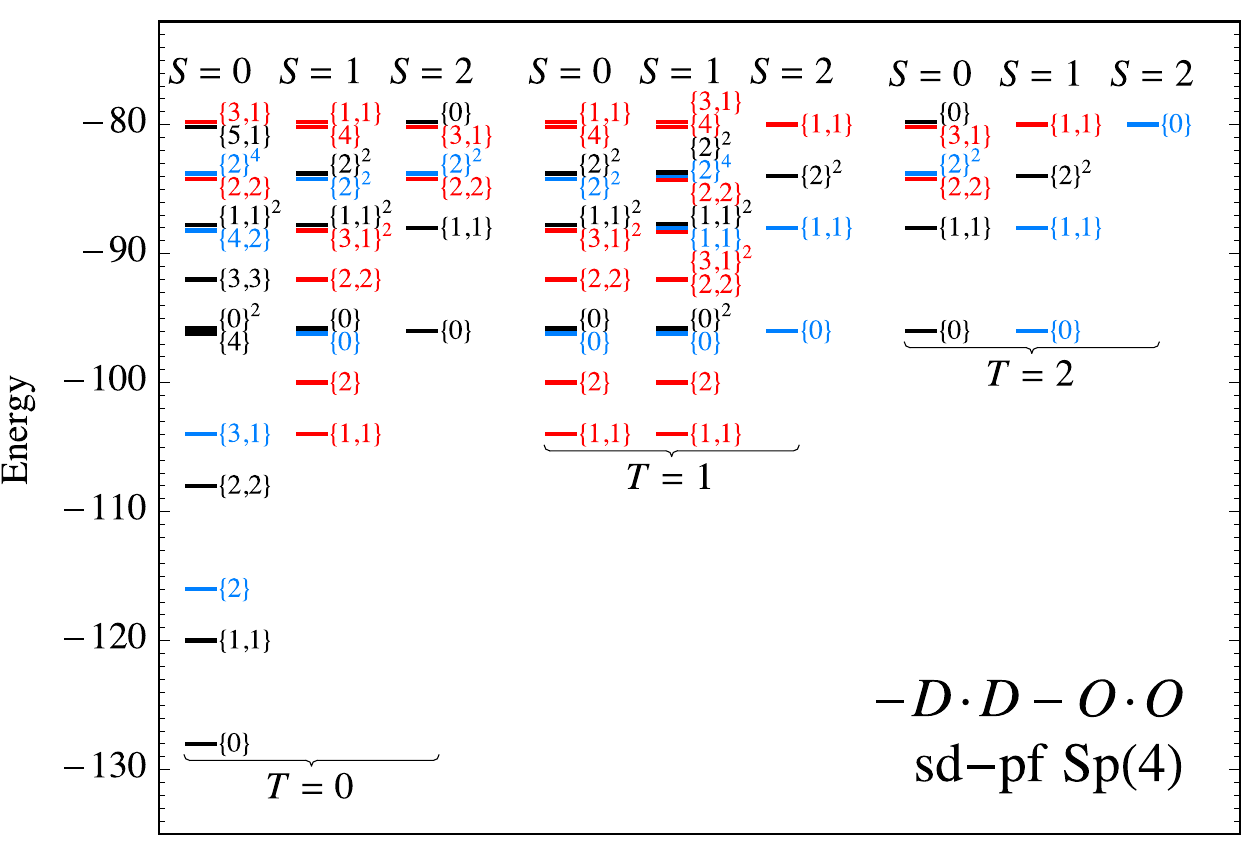}
\includegraphics[height=6.5cm]{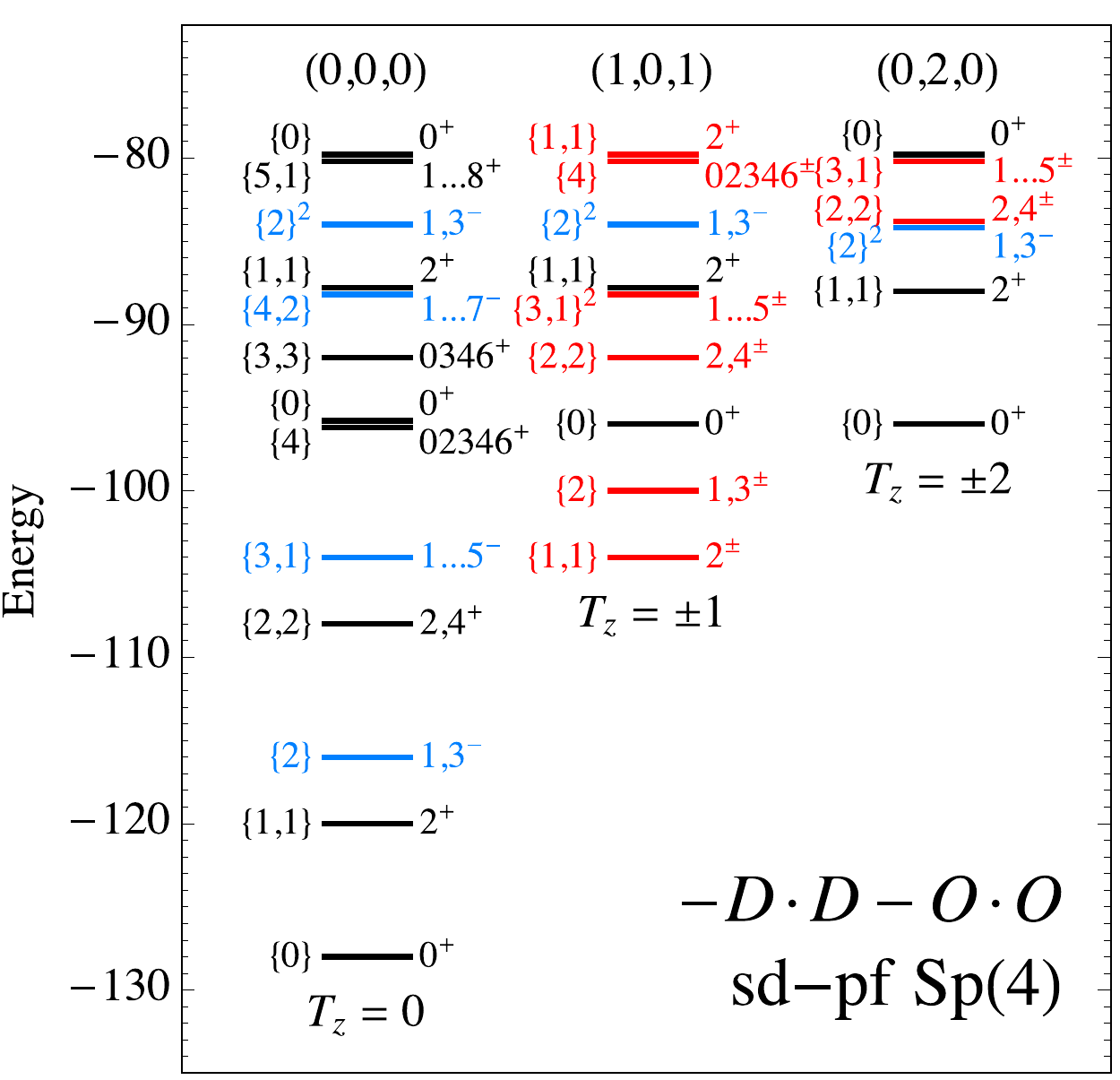}
\caption{Partial eigenspectrum of the hamiltonian
$-\hat D\cdot\hat D-\hat O\cdot\hat O$
for four nucleons in the $sd$--$pf$ shells.
Levels of positive (negative) parity are black (blue)
while parity doublets are red.
Left panel:
Levels are labelled by the Sp(4) quantum numbers $\{\nu_1,\nu_2\}$
and values of the total spin $S$ and the total isospin $T$ are also indicated.
If an irreducible representation $\{\nu_1,\nu_2\}$
occurs $k$ times for a given $S$ and $T$,
this is indicated by a superscript as $\{\nu_1,\nu_2\}^k$.
Right panel:
Only levels in the favoured supermultiplets are shown,
which are $(0,0,0)$ for $T_z=0$,
$(1,0,1)$ for $T_z=\pm1$
and $(0,2,0)$ for $T_z=\pm2$.
Levels are labelled
by the Sp(4) quantum numbers $\{\nu_1,\nu_2\}$ on the left
and by the orbital angular momenta $L$ and parity $\pi$ on the right.
The supermultiplet labels $(\lambda',\mu',\nu')$
and the isospin projection $T_z$ are also indicated,
and all levels have $S=0$.}
\label{f_octu2}
\end{figure*}
The basis states~(\ref{e_baso1}) in general do not carry a definite parity quantum number.
From the elementary relation for one particle,
\begin{equation*}
\hat P^{-1}a^\dag_{\ell m_\ell sm_stm_t}\hat P=(-)^\ell a^\dag_{\ell m_\ell sm_stm_t},
\end{equation*}
where $\hat P$ is the parity operator,
it follows that
\begin{equation*}
\hat P^{-1}\hat G^{(\lambda)}_\mu(\ell\ell')\hat P=(-)^{\ell+\ell'}\hat G^{(\lambda)}_\mu(\ell\ell'),
\end{equation*}
and therefore
\begin{equation*}
\hat P^{-1}\hat G^{(0\lambda)}_{0\mu}\hat P=\hat G^{(\lambda0)}_{\mu0},
\quad
\hat P^{-1}\hat G^{(\lambda0)}_{\mu0}\hat P=\hat G^{(0\lambda)}_{0\mu}.
\end{equation*}
This implies that the basis states~(\ref{e_baso1}) are transformed as
\begin{align*}
\lefteqn{\hat P
|[1^n];([\bar h];[\bar h_a]\langle\bar\omega_a\rangle\times[\bar h_b]\langle\bar\omega_b\rangle;
\langle\bar\omega_+\rangle L)\times[\bar h']ST\rangle}
%\\={}&\varphi(\bar\omega_a,\bar\omega_b,\bar\omega_+)
\\&=
\varphi
|[1^n];([\bar h];[\bar h_b]\langle\bar\omega_b\rangle\times[\bar h_a]\langle\bar\omega_a\rangle;
\langle\bar\omega_+\rangle L)\times[\bar h']ST\rangle,
\end{align*}
where $\varphi$ is a phase.
%which depends on $\bar\omega_a$, $\bar\omega_b$, and $\bar\omega_+$.
So one encounters the seemingly paradoxical situation
that the hamiltonian~(\ref{e_hamo1}), which does conserve parity,
has eigenstates~(\ref{e_baso1})
that in general do not carry the parity quantum number.
However, according to the energy formula~(\ref{e_eigo1})
the states connected by the parity transformation,
that is, states with the indices `$a$' and `$b$' interchanged,
%$\langle\bar\omega_a\rangle\times\langle\bar\omega_b\rangle$
%and $\langle\bar\omega_b\rangle\times\langle\bar\omega_a\rangle$,
are degenerate in energy.
Consequently, the states
\begin{align*}
\lefteqn{|[1^n];([\bar h];[\bar h_a]\langle\bar\omega_a\rangle\times[\bar h_b]\langle\bar\omega_b\rangle;
\langle\bar\omega_+\rangle L)\times[\bar h']ST\rangle_\pm}
\\&\equiv
\frac{1}{\sqrt2}\left(1\pm\hat P\right)
\\&\phantom{=}\times
|[1^n];([\bar h];[\bar h_a]\langle\bar\omega_a\rangle\times[\bar h_b]\langle\bar\omega_b\rangle;
\langle\bar\omega_+\rangle L)\times[\bar h']ST\rangle,
\end{align*}
are also eigenstates of the hamiltonian~(\ref{e_hamo1})
and they do carry a definite parity quantum number,
\begin{align}
\lefteqn{\hat P|[1^n];([\bar h];[\bar h_a]\langle\bar\omega_a\rangle\times[\bar h_b]\langle\bar\omega_b\rangle;
\langle\bar\omega_+\rangle L)\times[\bar h']ST\rangle_\pm}
\nonumber\\&=
\pm|[1^n];([\bar h];[\bar h_a]\langle\bar\omega_a\rangle\times[\bar h_b]\langle\bar\omega_b\rangle;
\langle\bar\omega_+\rangle L)\times[\bar h']ST\rangle_\pm.
\nonumber
\end{align}
These are the parity doublets
that occur for a reflection-asymmetric quantum-mechanical system.

The eigenspectrum of the hamiltonian~(\ref{e_hamo1}),
including the parity of the levels, can now be determined.
Figure~\ref{f_octu2} shows the example of four nucleons in the $sd$--$pf$ shells,
in which case the hamiltonian reduces to \mbox{$-\hat D\cdot\hat D-\hat O\cdot\hat O$}
and the symmetry is Sp(4).
The spectrum contains states~(\ref{e_baso1})
with identical indices `$a$' and `$b$',
$[\bar h_a]\{\bar\nu_a\}=[\bar h_b]\{\bar\nu_b\}$,
which have either positive or negative parity
(black and blue, respectively, in figure~\ref{f_octu2}).
Furthermore, levels with $[\bar h_a]\{\bar\nu_a\}\neq [\bar h_b]\{\bar\nu_b\}$
correspond to parity doublets (shown in red).
The states shown in the left panel of figure~\ref{f_octu2}
belong to all five possible supermultiplets
and the spectrum is complete up to an energy $E=-80$.
Retaining only the states contained in the favoured supermultiplets,
we find the spectra shown in the right panel of figure~\ref{f_octu2}
for $T_z=0$, $\pm1$ and $\pm2$.
A striking feature of the eigenspectrum of the hamiltonian~(\ref{e_hamo1})
is the presence of several parity doublets at low excitation energy
in the odd--odd system with $T_z=\pm1$.

\subsection{The limit of large oscillator shells}
\label{ss_largeN}
While the components in equation~(\ref{e_ang1}) are associated with the orbital angular momentum,
it is not {\it a priori} obvious that $\hat D_\mu$ and $\hat O_\mu$
have anything to do with the physical dipole and octupole operators.
This relation is discussed in this subsection.

In second quantization any SU(4)-scalar operator of multipolarity $\lambda$
can be written as
\begin{equation}
\hat{\cal T}^{(\lambda)}_\mu=
\sum_{\ell\ell'}
t^{(\lambda)}_{\ell\ell'}({\cal T})\hat G^{(\lambda)}_\mu(\ell\ell'),
\label{e_mult}
\end{equation}
where the coefficients $t^{(\lambda)}_{\ell\ell'}({\cal T})$
can be considered as the definition of the operator $\hat{\cal T}^{(\lambda)}_\mu$.
In second quantization
the multipole operator $r^\lambda Y_{\lambda\mu}$ reads
\begin{equation}
\hat T_\mu(r^\lambda Y_\lambda)=-\sqrt\frac{128\pi}{N+1}\sum_{\ell\ell'}
\frac{\langle\ell||r^\lambda Y_\lambda||\ell'\rangle}{\sqrt{2\lambda+1}}
\hat G^{(\lambda)}_\mu(\ell\ell'),
\label{e_dipoct1}
\end{equation}
where $-\sqrt{128\pi/(N+1)}$ is a conventional factor
introduced for reasons explained below.
The choice of the potential well
determines the coefficients $t^{(\lambda)}_{\ell\ell'}(r^\lambda Y_\lambda)$
since the reduced matrix elements $\langle\ell||r^\lambda Y_\lambda||\ell'\rangle$
depend on radial integrals.
These are well known for the harmonic oscillator~\cite{Talmi93},
leading to the coefficients
\begin{equation*}
t^{(\lambda)}_{\ell\ell'}{(r^\lambda Y_\lambda)}=
-(-)^\ell
\sqrt{\frac{32}{N+1}}\hat\ell\hat\ell'
\left(\!\!\begin{array}{ccc}
\ell&\lambda&\ell'\\
0&0&0
\end{array}\!\!\right)
I_{n\ell n'\ell'}^\lambda,
\end{equation*}
where $I_{n\ell n'\ell'}^\lambda$ is the radial integral
\begin{equation*}
I_{n\ell n'\ell'}^\lambda=
\int_0^{+\infty}r^\lambda R_{n\ell}(r)R_{n'\ell'}(r)r^2dr.
\end{equation*}
Alternatively, one may wish to consider the multipole operator $Y_{\lambda\mu}$,
which in second quantization has the coefficients
\begin{equation*}
t^{(\lambda)}_{\ell\ell'}{(Y_\lambda)}=
-(-)^\ell
\sqrt{\frac{32}{N+1}}\hat\ell\hat\ell'
\left(\!\!\begin{array}{ccc}
\ell&\lambda&\ell'\\
0&0&0
\end{array}\!\!\right),
\end{equation*}
which is identical to the expression for $t^{(\lambda)}_{\ell\ell'}(r^\lambda Y_\lambda)$
but for the radial integral $I_{n\ell n'\ell'}^\lambda$.

As shown in section~\ref{s_elliott},
Elliott's SU(3) model arises
because the components of the orbital angular momentum operator~(\ref{e_ang}),
together with those of the quadrupole operator~(\ref{e_quad}),
close under commutation.
No closure property is exactly valid
for $\hat T_\mu(r^\lambda Y_\lambda)$ if $\lambda\neq2$
but an approximate treatment is possible in the limit of large oscillator shells,
$N\rightarrow\infty$.
To see this point,
we introduce into the generators~(\ref{e_sopm2}) of ${\rm SOp}_-(\Omega)$
the explicit expressions~(\ref{e_dtensor2}) for the double tensors 
$\hat G^{(0\lambda)}_{0\mu}$ and $\hat G^{(\lambda0)}_{\mu0}$,
leading to the coefficients
\begin{equation*}
t^{(\lambda)}_{\ell\ell'}(T^{(\lambda)})=
-(-)^{\ell+N}
\sqrt{32}\;\hat\ell\hat\ell'
\left\{\!\!\begin{array}{ccc}
\ell&\lambda&\ell'\\
{\frac12}N&{\frac12}N&{\frac12}N
\end{array}\!\!\right\}.
\end{equation*}
Since in the limit of large $N$ one has the property~\cite{Ponzano68}
\begin{equation*}
\lim_{N\rightarrow\infty}
(-)^N\sqrt{N+1}
\left\{\!\!\begin{array}{ccc}
\ell&\lambda&\ell'\\
{\frac12}N&{\frac12}N&{\frac12}N
\end{array}\!\!\right\}=
\left(\!\!\begin{array}{ccc}
\ell&\lambda&\ell'\\
0&0&0
\end{array}\!\!\right),
\end{equation*}
we conclude that the matrix elements
of the generators $\hat T^{(\lambda)}_\mu$ of ${\rm SOp}_-(\Omega)$
tend to those of the operators $Y_{\lambda\mu}$,
\begin{equation*}
\lim_{N\rightarrow\infty}t^{(\lambda)}_{\ell\ell'}(T^{(\lambda)})=
t^{(\lambda)}_{\ell\ell'}{(Y_\lambda)}.
\end{equation*}
This approximation is reasonable even for relatively low values of $N$
as long as $\ell$ and $\ell'$ are not too large.
The factor $r^\lambda$, however, introduces a radial dependence
(contained in the integrals $I_{n\ell n'\ell'}^\lambda$)
which is not included in the algebraic definition of the generators.

\begin{table*}
\centering
\caption{Coefficients $t^{(\lambda)}_{\ell\ell'}$ for dipole and octupole operators
appropriate for the $sd$--$pf$ and $pf$--$sdg$ shells.}
\label{t_coefs}
\smallskip
\begin{tabular}{cccccccccc}
\hline\hline
$N$&shells&operator&
$t^{(1)}_{sp}$&$t^{(3)}_{sf}$&$t^{(\lambda)}_{pd}$&$t^{(3)}_{pg}$&$t^{(\lambda)}_{df}$&$t^{(\lambda)}_{fg}$\\
\hline
3&$sd$-$pf$&$\hat T^{(1)}_\mu$&$2\sqrt{2}$&---&$8\sqrt{\frac{1}{5}}$&---&$2\sqrt{\frac{14}{5}}$&---\\
&&$Y_{1\mu}$&$2\sqrt{2}$&---&$4$&---&$2\sqrt{6}$&---\\
&&$rY_{1\mu}$&$2\sqrt{5}$&---&$-4$&---&$2\sqrt{21}$&---\\
\hline
&&$\hat T^{(3)}_\mu$&---&$2\sqrt{2}$&$-2\sqrt{\frac{6}{5}}$&---&$4\sqrt{\frac{6}{5}}$&---\\
&&$Y_{3\mu}$&---&$2\sqrt{2}$&$-6\sqrt{\frac{2}{7}}$&---&$-4\sqrt{\frac{2}{3}}$&---\\
&&$r^3Y_{3\mu}$&---&$-3\sqrt{70}$&$6\sqrt{14}$&---&$-6\sqrt{21}$&---\\
\hline
4&$pf$-$sdg$&$\hat T^{(1)}_\mu$&
$4\sqrt{\frac{2}{5}}$&---&$2\sqrt{\frac{14}{5}}$&---&$8\sqrt{\frac{1}{5}}$&$4\sqrt{\frac{3}{5}}$\\
&&$Y_{1\mu}$&$4\sqrt{\frac{2}{5}}$&---&$8\sqrt{\frac{1}{5}}$&---&$4\sqrt{\frac{6}{5}}$&$8\sqrt{\frac{2}{5}}$\\
&&$rY_{1\mu}$&$-8\sqrt{\frac{1}{5}}$&---&$4\sqrt{\frac{14}{5}}$&---&$-4\sqrt{\frac{6}{5}}$&$24\sqrt{\frac{1}{5}}$\\
\hline
&&$\hat T^{(3)}_\mu$&---&$4\sqrt{\frac{2}{5}}$&$-8\sqrt{\frac{3}{35}}$&$12\sqrt{\frac{1}{35}}$
&$4\sqrt{\frac{3}{35}}$&$6\sqrt{\frac{22}{35}}$\\
&&$Y_{3\mu}$&---&$4\sqrt{\frac{2}{5}}$&$-12\sqrt{\frac{2}{35}}$&$8\sqrt{\frac{6}{35}}$
&$-8\sqrt{\frac{2}{15}}$&$-24\sqrt{\frac{1}{55}}$\\
&&$r^3Y_{3\mu}$&---&$12\sqrt{\frac{14}{5}}$&$-78\sqrt{\frac{1}{5}}$&$-36\sqrt{\frac{6}{5}}$
&$24\sqrt{\frac{6}{5}}$&$-18\sqrt{\frac{22}{5}}$\\
\hline\hline
\end{tabular}
\end{table*}
In table~\ref{t_coefs} are given the coefficients $t^{(\lambda)}_{\ell\ell'}$
for three different kinds of dipole and octupole operators
appropriate for the $sd$--$pf$ and $pf$--$sdg$ shells.
The first line of each entry lists the coefficients
for the generators $\hat T^{(\lambda)}_\mu$ defined in equation~(\ref{e_sopm2}).
The second line gives the coefficients
appropriate for the dipole or octupole operators $Y_{\lambda\mu}$.
The normalization $-\sqrt{32/(N+1)}$ is chosen such that the first coefficient,
$t^{(1)}_{sp}$ or $t^{(3)}_{sf}$, coincides with the algebraic definition in the first line.
As shown above,
the matrix elements of $Y_{\lambda\mu}$
tend to those of the generators $\hat T^{(\lambda)}_\mu$,
in the limit of large $N\gg\ell,\ell'$.
It is indeed seen that deviations grow with increasing $\ell$ and $\ell'$,
especially for the octupole operator.
The last line of each entry lists the coefficients
appropriate for the dipole or octupole operators $r^\lambda Y_{\lambda\mu}$
and calculated with harmonic-oscillator radial wave functions.
Since the same normalization factor %$-\sqrt{128\pi/(N+1)}$
is adopted as for $Y_{\lambda\mu}$,
the first coefficient, $t^{(1)}_{sp}$ or $t^{(3)}_{sf}$, deviates from the algebraic definition
and this deviation depends on the radial integrals $I_{n\ell n'\ell'}^\lambda$.

We conclude this subsection with two further remarks
concerning the results in table~\ref{t_coefs}.
The first concerns the signs of the coefficients
$t^{(\lambda)}_{\ell\ell'}{(Y_\lambda)}$
and $t^{(\lambda)}_{\ell\ell'}{(r^\lambda Y_\lambda)}$
which are seen in some cases to deviate
from those of the coefficients of $\hat T^{(\lambda)}_\mu$.
This can be remedied by a change of phase of the type
\begin{equation*}
a^\dag_{\ell st}\mapsto -a^\dag_{\ell st},
\end{equation*}
for certain single-particle orbital angular momenta $\ell$.
Closure properties are not affected by such changes of phase
and, consequently, the solvability property proven above remains valid.
This property is well known for bosons~\cite{Isacker85,Shirokov98}
and also applies to fermionic systems.
For the $sd$--$pf$ and $pf$--$sdg$ shells enough freedom exists to accommodate
any combination of signs of the coefficients $t^{(\lambda)}_{\ell\ell'}$.
In other words, there always exists a change of phase of the $a^\dag_{\ell_ist}$
for certain single-particle orbital angular momenta $\ell_i\in\{0,1,\dots,N\}$
such that the signs of $t^{(\lambda)}_{\ell\ell'}$ in the generators $\hat T^{(\lambda)}_\mu$
are as required.
This remains true for the dipole operator for any $N$
but is no longer generally valid for the octupole operator if $N\geq5$,
that is, for the $sdg$--$pfh$ shells and beyond.

The second remark concerns the dependence
of the coefficients $t^{(\lambda)}_{\ell\ell'}{(r^\lambda Y_\lambda)}$
on the radial integrals $I_{n\ell n'\ell'}^\lambda$.
The coefficients listed in table~\ref{t_coefs} are for a harmonic oscillator
and other potentials lead to different results.
Suppose we can find a potential with the property
\begin{equation*}
I_{n\ell n'\ell'}^\lambda=
c\frac{\left\{\!\!\begin{array}{ccc}
\ell&\lambda&\ell'\\
{\frac12}N&{\frac12}N&{\frac12}N
\end{array}\!\!\right\}}
{\left(\!\!\begin{array}{ccc}
\ell&\lambda&\ell'\\
0&0&0
\end{array}\!\!\right)},
\end{equation*}
(where $c$ is an arbitrary constant)
for all $(n,\ell,n',\ell')$ such that $2n+\ell=2n'+\ell'-1=N$.
For such a potential
the following proportionality is valid without any approximation:
\begin{equation*}
t^{(\lambda)}_{\ell\ell'}(T^{(\lambda)})\propto
t^{(\lambda)}_{\ell\ell'}{(r^\lambda Y_\lambda)},
\end{equation*}
implying an {\em exact} algebraic realization of the operator $r^\lambda Y_\lambda$.

\subsection{The octupole hamiltonian}
\label{ss_octu}
The hamiltonian~(\ref{e_hamo1}) contains the dipole interaction $\hat D\cdot\hat D$,
which we may wish to eliminate.
In particular, for $N=3$ and $N=4$ (the $sd$--$pf$ and $pf$--$sdg$ shells),
two cases of main interest,
elimination of $\hat D\cdot\hat D$ leads to a pure octupole interaction $\hat O\cdot\hat O$.
This elimination is achieved by noting that,
since $\hat L_\mu=\hat L_{a\mu}+\hat L_{b\mu}$
and $\hat D_\mu=\hat L_{a\mu}-\hat L_{b\mu}$,
one has
\begin{equation}
\hat D\cdot\hat D=2\hat L_a\cdot\hat L_a+2\hat L_b\cdot\hat L_b-\hat L\cdot\hat L,
\label{e_dipole}
\end{equation}
and therefore the dipole interaction is diagonal in the basis
\begin{align}
&\begin{array}{cccccccccc}
{\rm U}(\Omega^2)&\!\!\!\!\supset\!\!\!\!&
{\rm U}_a(\Omega)&\!\!\!\!\otimes\!\!\!\!&{\rm U}_b(\Omega)&\!\!\!\!\supset\!\!\!\!&
{\rm SOp}_a(\Omega)&\!\!\!\!\otimes\!\!\!\!&{\rm SOp}_b(\Omega)\\
\downarrow&&\downarrow&&\downarrow&&\downarrow&&\downarrow\\[0mm]
[\bar h]&&[\bar h_a]&&[\bar h_b]&&\langle\bar\omega_a\rangle&&\langle\bar\omega_b\rangle
\end{array}
\nonumber\\\nonumber\\[-10pt]
&\begin{array}{cccccccccc}
\supset\!\!\!\!&{\rm SO}_a(3)&\!\!\!\!\otimes\!\!\!\!&{\rm SO}_b(3)&\!\!\!\!\supset\!\!\!\!&{\rm SO}_+(3)\\
&\downarrow&&\downarrow&&\downarrow&\\[0mm]
&L_a&&L_b&&L
\end{array},
\label{e_claso2}
\end{align}
with eigenstates
\begin{equation*}
|[1^n];([\bar h];[\bar h_a]\langle\bar\omega_a\rangle L_a\times[\bar h_b]\langle\bar\omega_b\rangle L_b;
L)\times[\bar h']ST\rangle,
\end{equation*}
and with energy eigenvalues
\begin{equation*}
2L_a(L_a+1)+2L_b(L_b+1)-L(L+1).
\end{equation*}
The bases~(\ref{e_claso1}) and~(\ref{e_claso2})
are connected by a unitary transformation,
\begin{align*}
\lefteqn{|\langle\bar\omega_a\rangle\times\langle\bar\omega_b\rangle;\langle\bar\omega_+\rangle L\rangle}
\\&=
\sum_{L_aL_b}
\left\langle\begin{array}{cc|c}
\langle\bar\omega_a\rangle&\langle\bar\omega_b\rangle&\langle\bar\omega_+\rangle\\
L_a&L_b&L
\end{array}\right\rangle
|\langle\bar\omega_a\rangle L_a\times\langle\bar\omega_b\rangle L_b;L\rangle,
\end{align*}
where the symbol in angle brackets is an isoscalar factor~\cite{Wybourne74},
associated with either ${\rm Sp}(\Omega)\supset{\rm SO}(3)$ if $\Omega$ is even
or ${\rm SO}(\Omega)\supset{\rm SO}(3)$ if $\Omega$ is odd.
Note that this transformation does not depend
on other labels of the states~(\ref{e_claso1}) and~(\ref{e_claso2}),
which are therefore suppressed.
The combination of the previous results
leads to the following expression for the matrix elements of $\hat D\cdot\hat D$:
\begin{align}
\lefteqn{\langle\langle\bar\omega_a\rangle\times\langle\bar\omega_b\rangle;\langle\bar\omega_+\rangle L|
\hat D\cdot\hat D
|\langle\bar\omega_a\rangle\times\langle\bar\omega_b\rangle;\langle\bar\omega'_+\rangle L\rangle}
\nonumber\\&=
\sum_{L_aL_b}
[2L_a(L_a+1)+2L_b(L_b+1)-L(L+1)]
\nonumber\\&\phantom{=}\times
\left\langle\begin{array}{cc|c}
\langle\bar\omega_a\rangle&\langle\bar\omega_b\rangle&\langle\bar\omega_+\rangle\\
L_a&L_b&L
\end{array}\right\rangle
\left\langle\begin{array}{cc|c}
\langle\bar\omega_a\rangle&\langle\bar\omega_b\rangle&\langle\bar\omega'_+\rangle\\
L_a&L_b&L
\end{array}\right\rangle.
\nonumber
\end{align}
The octupole interaction $\hat O\cdot\hat O$ is not diagonal in the basis~(\ref{e_baso1})
but rather block diagonal.
The blocks consist of basis states
that have different labels $\langle\bar\omega_+\rangle$
but otherwise identical quantum numbers.
Provided the necessary isoscalar factors are known,
the determination of the eigenspectrum of $\hat O\cdot\hat O$
only requires the diagonalization of matrices of modest size (dimension $\sim$10).

\section{Concluding remarks}
\label{s_conc}
We have discussed in this paper
two analytic solutions of the spherical shell model
that make contact with the geometric collective model.
The first is based on Elliott's well-known SU(3) symmetry,
which is generated by a spin--isospin-scalar quadrupole interaction $r^2Y_2\cdot r^2Y_2$
and provides a natural explanation of the phenomenon of nuclear rotation.
A prerequisite for its existence
is the spin--isospin SU(4) symmetry of Wigner's supermultiplet model.
Only when the condition of exact solvability is relaxed
can one propose extensions such as pseudo-SU(3) or quasi-SU(3)
that apply to more realistic $jj$-coupled situations.

We have shown in this paper
that another analytic solution of the spherical shell model exists,
generated by a spin--isospin-scalar octupole interaction,
which in the limit of large oscillator shells tends to its geometric equivalent $Y_3\cdot Y_3$.
A natural outcome of the ensuing symmetry,
which can be either orthogonal or symplectic,
is the presence in the excitation spectrum of parity doublets,
as is required of a reflection-asymmetric quantum-mechanical object.

The octupole solution of the spherical shell model presented in this paper
is developed to a schematic level only
and it is too early to tell
whether it can be extended to more realistic situations
with applications to actual nuclei.
Among the open problems that need to be explored
are the non-degeneracy of the lower and upper oscillator shells,
and the departure from spin--isospin symmetry
({\it e.g.}, the spin--orbit interaction).
Furthermore, in the heavy regions of the periodic table,
where octupole deformation is relevant,
the neutron and proton orbitals are different,
though for both it is possible to choose neighboring shells with different parities.
Also, the questions of possible effects
resulting from the differences
between the algebraic and geometric octupole operators
as well as the elimination of states due to spurious centre-of-mass motion
need to be addressed.

The existence of an octupole solution of the spherical shell model,
as an alternative to Elliott's quadrupole solution,
also raises the question whether the two deformations
can be combined into one model.
It may not be possible to elaborate
a fully solvable quadrupole--octupole shell-model hamiltonian
but recent studies indicate
that the application range of dynamical symmetries
is considerably extended through partial solvability~\cite{Leviatan11}.
It will therefore be of interest to apply this notion
to the two deformed symmetries of the spherical shell model
discussed in this paper.

\begin{ack}
This work was partially supported (SP) by FUSTIPEN
(French-US Theory Institute for Physics with Exotic Nuclei)
under DOE grant DE-FG02-10ER41700.
\end{ack}

\section*{Appendix: Branching rules}
The applications presented in this paper
require the knowledge of a variety of branching rules,
the problem of which can be formulated as follows.
Given two algebras $G_1$ and $G_2$ with $G_1\supset G_2$,
what irreducible representations of $G_2$
are contained in a given irreducible representation of $G_1$?
Note that, if $G_2={\rm SO}(3)$,
this amounts to finding the (orbital) angular momentum content
of a given irreducible representation of $G_1$.

The branching rules associated with unitary algebras
concern the following cases:
\begin{itemize}
\item
${\rm U}(n_1)\supset{\rm U}(n_2),\quad n_1>n_2$,
\item
${\rm U}(n_1n_2)\supset{\rm U}(n_1)\otimes{\rm U}(n_2)$,
\item
${\rm U}(n_1+n_2)\supset{\rm U}(n_1)\otimes{\rm U}(n_2)$.
\end{itemize} 
These, of course, are known since long from the classical theory
of group characters and representations,
as described, for example, in the monographs
by Murnaghan~\cite{Murnaghan38} and Littlewood~\cite{Littlewood40}.
The purpose of this appendix is to show
that all such branching rules can be derived
from a few known results concerning Young diagrams
that can be programmed in a symbolic language
like {\tt Mathematica}~\cite{Isackerun2}.

The first result concerns the (outer) multiplication
of two Young diagrams $[\bar h_1]$ and $[\bar h_2]$,
which can be written as
\begin{equation*}
[\bar h_1]\times[\bar h_2]=
\sum_{\bar h}\Gamma_{\bar h_1\bar h_2\bar h}[\bar h],
\end{equation*}
where $\Gamma_{\bar h_1\bar h_2\bar h}$ denotes the number of times ($0,1,2,\dots$)
$[\bar h]$ occurs in the outer product $[\bar h_1]\times[\bar h_2]$.
The rules for calculating such outer products of Young diagrams
can be found in many textbooks (see, for example, section~7-12 of Ref.~\cite{Hamermesh62})
and will not be repeated here.

The second result states that any Young diagram
can be written as a linear combination of products of symmetric Young diagrams.
The explicit decomposition for a Young diagram of length $s$,
$[\bar h]\equiv[h_1,h_2,\dots,h_s]$, reads
\begin{equation*}
[\bar h]=
\left|\begin{array}{llll}
[h_1]&[h_1+1]&\cdots&[h_1+s-1]\\[0pt]
[h_2-1]&[h_2]&\cdots&[h_2+s-2]\\[0pt]
\vdots&\vdots&\ddots&\vdots\\[0pt]
[h_s-s+1]&[h_s-s+2]&\cdots&[h_s]
\end{array}\right|,
\end{equation*}
with the convention
that $[h]=1$ for $h=0$ and $[h]=0$ for $h<0$.
The most elementary example is the decomposition of $[1,1]$,
for which $h_1=h_2=1$ and $s=2$, and therefore
\begin{equation*}
[1,1]=
\left|\begin{array}{cc}[1]&[2]\\1&[1]\end{array}\right|=
[1]\times[1]-[2].
\end{equation*}
In general, we write the decomposition of $[\bar h]$ as 
\begin{equation*}
[\bar h]=\sum_{\{r_1,r_2,\dots,r_k\}}
a^{[\bar h]}_{r_1r_2\dots r_k}
\prod_{i=1}^k[r_i],
%[r_1]\times[r_2]\times\cdots\times[r_k],
\end{equation*}
with products $[r_1]\times[r_2]\times\cdots\times[r_k]$
of one-rowed Young diagrams of lengths $r_1,r_2,\dots,r_k$
and coefficients $a^{[\bar h]}_{r_1r_2\dots r_k}$
that are readily obtained from the determinantal expression.

To obtain branching rules
associated with the reduction $G_1={\rm U}(n)\supset G_2$,
one needs to specify what is the branching rule
for the fundamental representation $[1]$ of ${\rm U}(n)$.
This rule can be written generically as $[1]\mapsto[\bar f]$,
where $[\bar f]$ is an irreducible representation of $G_2$.
For example, in the reduction ${\rm U}(\Gamma)\supset{\rm U}(3)$
of Elliott's model for the oscillator shell with major quantum number $N$
[and $\Gamma\equiv(N+1)(N+2)/2$],
the fundamental representation $[1]$ of ${\rm U}(\Gamma)$
reduces to $[N,0,0]$ of U(3).
In this case therefore $[\bar f]=[N,0,0]$,
which expresses the fact that one nucleon in the $N$-shell
corresponds to $N$ oscillator quanta.
Given the property $[1]\mapsto[\bar f]$ in the reduction ${\rm U}(n)\supset G_2$,
the branching rule for a general irreducible representation $[\bar h]$ of ${\rm U}(n)$
can be written as
\begin{equation*}
[\bar h]\mapsto[\bar f]\odot[\bar h],
\end{equation*}
where $\odot$ is the operation of `plethysm',
whose properties are discussed below.
On the left-hand side of $\mapsto$
stands the irreducible representation $[\bar h]$ of ${\rm U}(n)$
while on the right-hand side
stands a sum of irreducible representations of $G_2$.
The latter is found by carrying out the plethysm,
which can be done by applying the following rules:
\begin{align*}
[\bar f]\odot[1]&=[\bar f],
\\
[\bar f]\odot([\bar h_1]\pm[\bar h_2])&=[\bar f]\odot[\bar h_1]\pm[\bar f]\odot[\bar h_2],
\\
[\bar f]\odot([\bar h_1]\times[\bar h_2])&=\Bigl([\bar f]\odot[\bar h_1]\Bigr)\times\Bigl([\bar f]\odot[\bar h_2]\Bigr),
\end{align*}
where $\times$ refers to an outer multiplication in $G_2$.
The first rule is nothing but the definition of the branching rule for the fundamental representation
while the successive application of the second and third rules
lead to the result
\begin{equation*}
[\bar h]\mapsto\sum_{\{r_1,r_2,\dots,r_k\}}
a^{[\bar h]}_{r_1r_2\dots r_k}
\prod_{i=1}^k\bigl([\bar f]\odot[r_i]\bigr),
\end{equation*}
where the product refers to outer multiplication  in $G_2$.
This shows that the problem of working out the general plethysm $[\bar f]\odot[\bar h]$
is reduced to that of finding $[\bar f]\odot[h]$,
where the irreducible representation $[h]$ of ${\rm U}(n)$ is symmetric
and characterized by a one-rowed Young diagram.

For the branching rule associated with the reduction
${\rm U}(n_1n_2)\supset{\rm U}(n_1)\otimes{\rm U}(n_2)$
the following result is valid
for the symmetric representation $[h]$ of ${\rm U}(n_1n_2)$:
\begin{equation*}
[h]\mapsto\sum_{\bar s}[\bar s]\otimes[\bar s],
\end{equation*}
where the sum is over all Young diagrams with $h$ boxes
and at most $\min(n_1,n_2)$ rows,
$s_1+s_2+\cdots+s_k=h$ and $k\leq\min(n_1,n_2)$.
We do not give a formal proof of this result
but it is clear on intuitive grounds
since, for a symmetric wave function,
the symmetry character under the partial exchange of coordinates
must be compensated by an equivalent symmetry character
under the exchange of the remainder of the coordinates.

For the branching rule associated with the reduction
${\rm U}(n_1+n_2)\supset{\rm U}(n_1)\otimes{\rm U}(n_2)$
the following result is valid
for the symmetric representation $[h]$ of ${\rm U}(n_1+n_2)$:
\begin{equation*}
[h]\mapsto\sum_{r=0}^h[h-r]\otimes[r].
\end{equation*}
Again we do not give a formal proof of the result,
which in this case follows from the conservation of particle number.
Note that representations of ${\rm U}(n_1)\otimes{\rm U}(n_2)$
are written as $[\bar s_1]\otimes[\bar s_2]$
with the understanding that $[\bar s_i]$ is associated with ${\rm U}(n_i)$.

With these results the second and third of our list of unitary branching rules
are determined in complete generality
and the only remaining task is to find the branching rule
associated with ${\rm U}(n_1)\supset{\rm U}(n_2)$
for a symmetric representation $[h]$ of ${\rm U}(n_1)$.
Let us introduce the notation
\begin{equation*}
[\bar f]\odot[h]=\sum_{\bar s}x_{\bar s}[\bar s],
\end{equation*}
with coefficients $x_{\bar s}=0,1,2,\dots$ that are to be determined.
As can be understood from the general rules governing plethysms,
the sum is over all Young diagrams $[\bar s]$ with $fh$ boxes,
where $f$ is the number of boxes of the fundamental representation $[\bar f]$.
Furthermore, we denote by $[\bar s]/[1]$ all permissible Young diagrams
that can be obtained by deleting one box from $[\bar s]$ or, explicitly,
\begin{equation*}
[\bar s]/[1]\equiv\sum_{\bar u}\Gamma_{[1]\bar u\bar s}[\bar u].
\end{equation*}
Littlewood's third method~\cite{Littlewood40,Wybourne70}
states that the coefficients $x_{\bar s}$
are solutions of the equation
\begin{equation*}
\sum_{\bar s}x_{\bar s}\bigl([\bar s]/[1]\bigr)=
\bigl([\bar f]\odot[h-1]\bigr)\times\bigl([\bar f]/[1]\bigr),
\end{equation*}
which, if the fundamental representation has only one row, $[\bar f]=[f]$, reduces to
\begin{equation*}
\sum_{\bar s}x_{\bar s}\bigl([\bar s]/[1]\bigr)=
\bigl([f]\odot[h-1]\bigr)\times[f-1].
\end{equation*}
Since $[\bar f]\odot[1]=[\bar f]$,
the series on the right-hand side
can be generated by an induction hypothesis
and subsequent outer multiplications in ${\rm U}(n_2)$.
The coefficients $x_{\bar s}$ must be chosen
such that all terms on left- and right-hand sides are identical.
This condition is not always sufficient to determine all coefficients $x_{\bar s}$ uniquely
and must be supplemented by the equation
\begin{equation*}
\sum_{\bar s}d_{[\bar s]}(n_2)x_{[\bar s]}=d_{[h]}(n_1),
\end{equation*}
where $d_{[\bar s]}(n)$ is the dimension of the irreducible representation $[\bar s]$ in U($n$).
In all applications presented in this paper
we have found that the combination of the two conditions
suffices to determine the plethysm $[\bar f]\odot[h]$,
and therefore the general plethysm $[\bar f]\odot[\bar h]$
for the reduction ${\rm U}(n_1)\supset{\rm U}(n_2)$.

The branching rules associated with the restricted algebras
concern the following cases:
\begin{itemize}
\item
${\rm U}(n)\supset{\rm SO}(n)$,
\item
${\rm U}(n)\supset{\rm Sp}(n)$.
\end{itemize} 
These rules can be taken from Littlewood~\cite{Littlewood40}
and read in our notation
\begin{align*}
[\bar h]&\mapsto\langle\bar h\rangle+
\sum_{\bar\delta}\Gamma_{\bar\delta\bar\upsilon\bar h}\langle\bar\upsilon\rangle,
\\
[\bar h]&\mapsto\{\bar h\}+
\sum_{\bar\beta}\Gamma_{\bar\beta\bar\nu\bar h}\{\bar\nu\},
\end{align*}
for ${\rm U}(n)\supset{\rm SO}(n)$ and ${\rm U}(n)\supset{\rm Sp}(n)$, respectively.
The $\Gamma$s refer to outer multiplication in U($n$)
and the summations are over the partitions
\begin{align*}
[\bar\delta]&=[2],[4],[2^2],[6],[4,2],[2^3],\dots,
\\
[\bar\beta]&=[1^2],[2^2],[1^4],[3^2],[2^2,1^2],[1^6],\dots,
\end{align*}
that is, all partitions constructed out of $[2]$ and $[1^2]$, respectively.
Some of the irreducible representations of ${\rm SO}(n)$ and ${\rm Sp}(n)$
that appear in these branching rules have more labels than allowed
and must be eliminated or modified.
The appropriate modification rules are given by Wybourne~\cite{Wybourne70}.


\begin{thebibliography}{10}
\bibitem{RS80}
Ring~P and Schuck~P 1980
{\it The Nuclear Many-Body Problem}
(Springer, New York).

\bibitem{Bender03}
Bender~M, Heenen~P-H and Reinhard~P~G 2003
{\it Rev.\ Mod.\ Phys.\ } {\bf75} 121.

\bibitem{Bohr69}
Bohr~A and Mottelson~B~R 1969
{\it Nuclear Structure. I Single-Particle Motion}
(Benjamin, New York).

\bibitem{Mayer49}
Mayer~M~G 1949
{\it Phys.\ Rev.\ } {\bf75} 1969.

\bibitem{Jensen49}
Jensen~J~H~D, Suess~H and Haxel~O 1949
{\it Die Naturwissenschaften} {\bf36} 155.

\bibitem{Caurier05}
Caurier~E, Mart\' \i nez-Pinedo~G, Nowacki~F, Poves~A and Zuker~A~P 2005
{\it Rev.\ Mod.\ Phys.\ } {\bf77} 427.

\bibitem{Inglis53}
Inglis~D~R 1953
{\it Rev.\ Mod.\ Phys.\ } {\bf25} 390.

\bibitem{Wilkinson95}
Wilkinson~D~H 1995
{\it Annu.\ Rev.\ Nucl.\ Part.\ Sci.\ } {\bf45} 1.

\bibitem{Hammer13}
Hammer~H-W, Nogga~A and Schwenk~A 2013
{\it Rev.\ Mod.\ Phys.\ }{\bf85} 197.

\bibitem{Navratil07}
%Navr\'atil~P {\it et al.}
Navr\'atil~P, Gueorguiev~V~G, Vary~J~P, Ormand~W~E and Nogga~A 2007
{\it Phys.\ Rev.\ Lett.\ } {\bf99} 042501.

\bibitem{Otsuka10}
%Otsuka~T {\it et al.} 2010
Otsuka~T, Suzuki~T, Holt~J~D, Schwenk~A and Akaishi~Y 2010
{\it Phys.\ Rev.\ Lett.\ } {\bf105} 032501.

\bibitem{Casten85}
Casten~R~F 1985
{\it Nucl.\ Phys.\ A} {\bf443} 1.

\bibitem{Rayleigh79}
Lord Rayleigh 1879
{\it Proc.\ Roy.\ Soc.\ (London)} {\bf29} 71.

\bibitem{Weizsacker35}
von Weizs\"acker~C~F 1935
{\it Z.\ Phys.\ } {\bf96} 431.

\bibitem{Bohr37}
Bohr~N and Kalckar~F 1937
{\it Mat.\ Fys.\ Medd.\ Dan.\ Vid.\ Selsk.\ } {\bf14} no~10.

\bibitem {Feenberg39}
Feenberg~E 1939
{\it Phys.\ Rev.\ } {\bf55} 504.

\bibitem {Bohr39}
Bohr~N and Wheeler~J~A 1939 
{\it Phys. Rev.} {\bf56} 426.

\bibitem{Flugge41}
Fl\"ugge~S 1941
{\it Ann.\ Physik} {\bf39} 373.

\bibitem{Rainwater50}
Rainwater~J 1950
{\it Phys.\ Rev.\ } {\bf79} 432.

\bibitem{Bohr52}
Bohr~A 1952
{\it Mat.\ Fys.\ Medd.\ Dan.\ Vid.\ Selsk.\ } {\bf26} no~14.

\bibitem{Bohr53}
Bohr~A and Mottelson~B~R 1953
{\it Mat.\ Fys.\ Medd.\ Dan.\ Vid.\ Selsk.\  } {\bf27} no~16.

\bibitem{Eisenberg70}
Eisenberg~J~M and Greiner~W 1970
{\it Nuclear Models}
(North-Holland, Amsterdam).

\bibitem{Bohr75}
Bohr~A and Mottelson~B~R 1975
{\it Nuclear Structure. II Nuclear Deformations}
(Benjamin, New York).

\bibitem{Chacon76}
Chac\'on~E, Moshinsky~M and Sharp~R~T 1976
{\it J.\ Math.\ Phys.\ } {\bf17} 668.

\bibitem{Chacon77}
Chac\'on~E and Moshinsky~M 1977
{\it J.\ Math.\ Phys.\ } {\bf18} 870.

\bibitem{Gaffney13}
Gaffney~L~P {\it et al.} 2013
{\it Nature} {\bf497} 199.

\bibitem{Wigner37}
Wigner~E~P 1937
{\it Phys.\ Rev.\ } {\bf51} 106.

\bibitem{Talmi93}
Talmi~I 1993
{\it Simple Models of Complex Nuclei.
The Shell Model and Interacting Boson Model}
(Harwood Academic, Chur, Switzerland).

\bibitem{Bohm88}
Bohm~A, N\'eeman~Y, and Barut~A~O Eds 1988
{\it Dynamical Groups and Spectrum Generating Algebras} 
(World Scientific, Singapore).

\bibitem{Frank09}
Frank~A, Jolie~J and Van~Isacker~P 2009
{\it Symmetries in Atomic Nuclei.
From Isospin to Supersymmetry}
(Springer, New York).

\bibitem{Iachello06}
Iachello~F 2006
{\it Lie Algebras and Applications. Lecture Notes in Physics}
(Springer, Berlin).

\bibitem{Murnaghan38}
Murnaghan~F~D 1938
{\it The Theory of Group Representations}
(Dover, New York).

\bibitem{Littlewood40}
Littlewood~D~E 1940
{\it The Theory of Group Characters and Matrix Representations of Groups}
(Clarendon, Oxford).

\bibitem{Hamermesh62}
Hamermesh~M 1962
{\it Group Theory and Its Application to Physical Problems}
(Addison-Wesley, Reading MA).

\bibitem{Lipas93}
Lipas~P~O 1993
in {\it Algebraic Approaches to Nuclear Structure.
Interacting Boson and Fermion Models}, ed Casten~R~F
(Harwood Academic, Chur, Switzerland) p~47.

\bibitem{Isackerun1}
Van~Isacker~P,
{\tt Mathematica} program {\tt commutator.m}, unpublished.

\bibitem{Hecht69b}
Hecht~K~T and Pang~S~C 1969
{\it J.\ Math.\ Phys.\ } {\bf10} 1571.

\bibitem{Elliott81}
Elliott~J~P and J~A~Evans 1981
{\it Phys.\ Lett.\ B} {\bf101} 216.

\bibitem{Vogel00}
Vogel~P 2000
{\it Nucl.\ Phys.\ A} {\bf662} 148.

\bibitem{Frauendorf14}
Frauendorf~S and Macchiavelli~A~O 2014
{\it Progr.\ Part.\ Nucl.\ Phys.\ } {\bf78} 24.

\bibitem{Wigner37b}
Wigner~E~P 1937
{\it Phys.\ Rev.\ } {\bf51} 947.

\bibitem{Moller92}
M\"oller~P and Nix~R 1992
{\it Nucl.\ Phys.\ A} {\bf536} 20.

\bibitem{Satula97}
Satu\l a~W, Dean~D~J, Gary~J, Mizutori~S and Nazarewicz~W 1997
{\it Phys.\ Lett.\ B} {\bf407} 103.

\bibitem{Warner06}
Warner~D~D, Bentley~M~A and Van~Isacker~P 2006
{\it Nature Phys.\ } {\bf2} 311.

\bibitem{Elliott58}
Elliott~J~P 1958
{\it Proc.\ Roy.\ Soc.\ (London) A} {\bf245} 128 \& 562.

\bibitem{Jauch40}
Jauch~J~M and Hill~E~L 1940
{\it Phys.\ Rev.\ } {\bf57} 641.

\bibitem{Wybourne70}
Wybourne~B~G 1970
{\it Symmetry Principles and Atomic Spectroscopy}
(Wiley-Interscience, New York).

\bibitem{Zuker95}
Zuker~A~P, Retamosa~J, Poves~A and Caurier~E 1995
{\it Phys.\ Rev.\ C} {\bf52} R1741.

\bibitem{Zuker15} 
Zuker~A~P, Poves~A, Nowacki~F and Lenzi~S~M 2015
{\it Phys.\ Rev.\ C} {\bf92} 024320.

\bibitem{Hecht69}
Hecht~K~T and Adler~A 1969
{\it Nucl.\ Phys.\ A} {\bf137} 129.

\bibitem{Arima69}
Arima~A, Harvey~M and Shimizu~K 1969
{\it Phys.\ Lett.\ B} {\bf30} 517.

\bibitem{Ratna73}
Ratna Raju~R~D, Draayer~J~P and Hecht~K~T 1973
{\it Nucl.\ Phys.\ A} {\bf202} 433.

\bibitem{Ginocchio97}
Ginocchio~J~N 1997
{\it Phys.\ Rev.\ Lett.\ }{\bf78} 436.

\bibitem{Hirsch03}
Hirsch~J~G, Vargas~C~E, Popa~G and Draayer~J~P 2003
in {\it Computational and Group Theoretical Models in Nuclear Physics}
(World Scientific, Singapore) p~31.

\bibitem{NNDC}
http://www.nndc.bnl.gov/

\bibitem{Elliott55}
Elliott~J~P and Skyrme~T~H~R 1955
{\it Proc.\ Roy.\ Soc.\ (London) A} {\bf232} 561.

\bibitem{Cseh15}
Cseh~J 2015
{\it Phys.\ Lett.\ B} {\bf743} 213.

\bibitem{Rosensteel77}
Rosensteel~G and Rowe~D~J 1977
{\it Phys.\ Rev.\ Lett.\ } {\bf38} 10.

\bibitem{Rosensteel80}
Rosensteel~G and Rowe~D~J 1980
{\it Ann.\ Phys.\ (NY)} {\bf126} 343.

\bibitem{Gilmore74}
Gilmore~R 1974
{\it Lie Groups, Lie Algebras and Some of Their Applications}
(Wiley-Interscience, New York).

\bibitem{Draayer93}
Draayer~J~P 1993
in {\it Algebraic Approaches to Nuclear Structure.
Interacting Boson and Fermion Models},  ed Casten~R~F
(Harwood Academic, Chur, Switzerland) p~423.

\bibitem{Iachello87}
Iachello~F and Arima~A 1987
{\it The Interacting Boson Model}
(Cambridge University Press, Cambridge).

\bibitem{Engel85}
Engel~J and Iachello~F 1985
{\it Phys.\ Rev.\ Lett.\ } {\bf54} 1126.

\bibitem{Kusnezov89}
Kusnezov~D 1989
{\it J.\ Phys.\ A: Math.\ Gen.\ } {\bf22} 4271. 

\bibitem{Kusnezov90}
Kusnezov~D 1990
{\it J.\ Phys.\ A: Math.\ Gen.\ } {\bf23} 5673. 

\bibitem{Engel87}
Engel~J, Frank~A and Pittel~S 1987
{\it Phys.\ Rev.\ C }{\bf 35} 1973.

\bibitem{Kusnezov88}
Kusnezov~D~M 1988
{\it Nuclear Collective Quadrupole-Octupole Excitations
in the U(16) $spdf$ Interacting Boson Model},
Ph D thesis, Princeton University, unpublished.

\bibitem{Ponzano68}
Ponzano~G and Regge~T 1968
in {\it Spectroscopy and Group Theoretical Methods in Physics}
(North-Holland, Amsterdam) p~1.

\bibitem{Isacker85}
Van~Isacker~P, Frank~A and Dukelsky~J 1985
{\it Phys.\ Rev.\ C} {\bf31} 671.

\bibitem{Shirokov98}
Shirokov~A~M, Smirnova~N~A and Smirnov~Yu~F 1998
{\it Phys.\ Lett.\ B} {\bf434} 237.

\bibitem{Wybourne74}
Wybourne~B~G 1974
{\it Classical Groups for Physicists}
(Wiley-Interscience, New York).

\bibitem{Leviatan11}
Leviatan~A 2011 
{\it Prog.\ Part.\ Nucl.\ Phys.\ }{\bf66} 93.

\bibitem{Isackerun2}
Van~Isacker~P,
{\tt Mathematica} program {\tt young.m}, unpublished.

\end{thebibliography}
\end{document}